\documentclass[reprint,nofootinbib,amsmath,amssymb,aps,prd]{revtex4-1}

\usepackage{graphicx,dcolumn,bm,hyperref,float}

\newcommand{\be}{\begin{equation}}
\newcommand{\ee}{\end{equation}}
\newcommand{\bea}{\begin{eqnarray}}
\newcommand{\eea}{\end{eqnarray}}

\begin{document}

\title{Eternal Inflation and Reheating in the Presence of the Standard Model Higgs Field}

\author{Mudit Jain}
\email{mudit.jain@tufts.edu}
\affiliation{Institute of Cosmology, Department of Physics and Astronomy,
Tufts University, Medford, MA 02155, USA}

\author{Mark P. Hertzberg}
\email{mark.hertzberg@tufts.edu}
\affiliation{Institute of Cosmology, Department of Physics and Astronomy,
Tufts University, Medford, MA 02155, USA}

\begin{abstract}
We study the details of eternal inflation in the presence of a spectator Higgs field within the framework of the minimal Standard Model. We have recently shown that in the presence of scalar field(s) which allow inflation only within a finite domain of field values, the universe reaches a steady state where the normalized distribution for the field(s) converges to a steady state distribution~\cite{MM.1}. In this paper, we analyze this eternal inflation scenario with the renormalized Standard Model Higgs potential, since it also allows inflation in a finite domain, but turns over at high scales due to the running of the self-coupling, marking an exit from inflation. We compute the full steady state distribution for the Higgs using an integral evolution technique that we formulated in~\cite{MM.1} and the fractal dimension of the universe. We then obtain a  bound on the inflationary Hubble scale in order to have a large observable universe contained within the instability scale of $H\lesssim\mathcal{O}(10^9)$GeV depending on the top mass. Upon reheating of the universe, thermal fluctuations in the Higgs field could potentially pose another problem; however, we compute the rate of thermal bubble production and find that the probability of tunneling in the post-inflationary era is negligibly small even for very high reheat temperatures.
\end{abstract}

\maketitle

\section{\label{sec:intro}Introduction}

The discovery of the Higgs boson~\cite{Higgs.discovery}, which is required to maintain unitarity of W, Z boson scattering, completes the Standard Model (SM) of elementary particles. For the first time we now have a theory that does not violate tree-level unitarity. When combined with gravity, we know that the theory does violate unitarity at the Planck scale, but one wonders at what scale, if any, does new physics enter before the Planck scale. There are reasons to suspect all sorts of possible new physics associated with dark matter, grand unification, strong CP problem, hierarchy problem, baryogenesis, inflation, etc. However, with the lack of discovery of new physics at the LHC, we have to keep an open mind to the possibility that such new physics may only enter at extremely high energies. 

We can therefore enquire: what is the absolute highest scale at which new physics can enter? The Higgs sector itself may suggest a potential problem when studied beyond tree level: It is well known that when loop effects are taken into account, the Higgs self-coupling $\lambda$ becomes negative at energies around $10^{10}-10^{11}$\,GeV for the latest top mass bounds~\cite{Particle.data.latest} (mostly due to its interactions with top quarks and W, Z bosons).  This makes the renormalized Higgs potential turn over at some field value $h_{\text{ht}}$ and then become negative, meaning there is a potential instability at high field values (see ahead to Fig.~\ref{fig:runninglambda} (lower panel)). The lifetime of the electroweak vacuum is estimated to be much higher than the current lifetime of the universe~\cite{EWvevmetastbl.1,EWvevmetastbl.2,EWvevmetastbl.3,EWvevmetastbl.4}, and so it may not appear as an immediate problem. However, this instability may be very important in the very early universe when energy densities were extremely high and the Higgs field could explore large field values.

In particular, this instability has potentially huge ramifications for evolution during early universe inflation~\cite{A.Guth,A.Linde.1,A.Linde.2,A.Vilenkin}. This is because during inflation the Higgs field on super-horizon scales roughly undergoes quantum diffusion with step size $\sim H/(2\pi)$ per Hubble time, and drifts due to the renormalized potential $V$. For sufficiently large Hubble scale during inflation $H$, the quantum diffusion can lead to the Higgs field $h$ becoming larger than the instability scale $h_{\text{ht}}$. In this case the renormalized potential then causes the field to drift down the potential to even large field values, leading to a potential cosmic catastrophe and terminating inflation. In any case, it leads to field values on the ``wrong" side of the potential, which is evidently not our observed electroweak vacuum.

This leads to several important questions to be addressed: (i) How frequently does this allow for a large observable universe, that is at-least $\sim 60$ e-foldings across in each spatial direction? (ii) What bounds does this imply for the scale of inflation? (iii) What is the global structure of the universe in the context of eternal inflation? (iv) What are the effects of reheating? In this paper we will make progress in addressing these important questions.

Some of these questions have been the subject of considerable investigation in the literature, see Refs.~\cite{Espinosa.Riotto.1,Zurek.1,Zurek.2,Kohri.1,Kohri.2,Espinosa.Riotto.2,Figueroa:2015rqa,Ema:2016kpf,Enqvist:2016mqj,Gross:2015bea,Espinosa:2018eve,Joti:2017fwe,Ema:2016ehh,Markkanen:2018pdo,Saha:2016ozn,Espinosa:2018euj,Ema:2017rkk,Gong:2017mwt}. However, the existing works seem to be partially incomplete. For instance, usually a Gaussian ansatz is used in the estimations of the probability distribution of Higgs during inflation which is used to obtain bounds on inflationary Hubble by requiring a large observable universe ($\sim e^{180}\,H^{-3}$) to be born at the end. More importantly, it is often thought that the number of e-foldings provided by the inflaton (typically taken to be $N\approx 60$) is essential to estimate the fate of the observable universe, since otherwise for large enough e-foldings and high enough Hubble value the distribution grows indefinitely. However, as we will show in this paper, neither the Gaussian ansatz, nor the requirement of finite number of e-foldings (irrespective of inflationary scales), is needed to describe the physics. 

Indeed we will study the system much more systematically: we will consider the structure of the universe in the context of eternal inflation. This leads to results that are independent of the arbitrary choice of initial conditions for the Higgs. Other works in the literature often assume the Higgs begins near the origin and spreads out, then is truncated at $N\approx 60$. But as we will see, this is not necessary, as a steady state is established leading us to consider the possibility of much longer phases of inflation, including eternal inflation. The assumption of $N\approx 60$ is a conservative one and plausible. However, it is of significant interest to study the possibility that inflation lasted a lot longer and may very well have in fact been eternal. 
To our knowledge this is the first time such an investigation has taken place. We will show that the universe can obtain a steady state distribution and from this distribution we can determine the probability of obtaining a large observable universe. Along the way we will compute global properties, such as the fractal dimension of the universe. Furthermore, we will move beyond the Gaussian ansatz and carefully track the distribution of different Hubble patches more accurately than in the standard treatments.

At the time of reheating, temperature corrections to the Higgs potential can be important and potentially alter the constraints. For instance, in Ref.~\cite{Espinosa.Riotto.2} it is claimed that high scale reheating can save the universe from the instability; that high scale inflation is allowed, because even though the Higgs can end up on the wrong side of the potential during inflation, it is rescued during reheating due to thermal corrections to the potential. These claims are potentially problematic as it requires extraordinarily fast reheating to prevent the Higgs field from rolling down to a catastrophe, and such problems were not properly taken into account. At the other end of the spectrum, Refs.~\cite{Kohri.1,Kohri.2} claim that an upper bound on the reheat temperature can be obtained, even if the scale of inflation is relatively low, because they claim the large thermal fluctuations in the Higgs field can cause it to be thrown over to the wrong side, unless the temperature is sufficiently low. However, the authors only considered the Higgs distribution at a single point in space. Instead a correct analysis would be to compute the probability of bubble formation of the Higgs at finite temperature. We will do this and show that even for extremely high reheat temperatures, the tunneling probability is small enough for a true vacuum bubble to form in the observable universe. In this sense, finite temperature SM poses no problems for the observable universe.

In typical models of inflation, it is well known that the universe on large scales undergoes eternal inflation (see~\cite{A.Linde.1,A.Vilenkin}) and generates infinite numbers of patches in which inflation has ended. In this work we assume that inflation is driven by some other field in some potential, characterized by some scale $H$. In addition the Higgs field can act as a spectator field (to be clear, we are {\em not} assuming the ``Higgs-inflation" scenario often studied in the literature, where the Higgs drives inflation, especially because the latest top mass bounds suggests this is unlikely to be viable). Nevertheless the Higgs can act as a field to cause inflation to terminate. In the case of the SM Higgs, inflation is allowed to continue in regions that contain Higgs within its slow-roll end point $h_{\text{end}}$ (on the unstable side of its potential), while other regions exit inflation\footnote{Whether such regions form an AdS crunch or a black hole, etc, depends on the details of very high energy physics; and will not be pursued here.}. Then it is important to note that the distribution of Higgs in each Hubble patch ultimately converges to a constant steady state distribution~\cite{MM.1} and any initial transient behavior washes out. Therefore, there is no dependence on the amount of e-foldings that the inflaton may provide. All interesting statistical quantities, like the average sizes of regions containing Higgs within $h_{\text{ht}}$, the bound on $H$ to ensure the average size to be at least that of the observable universe, the fractal dimensions of the eternally inflating universe, etc, must be computed at steady state. Finally, in order for our observable universe to be achieved, inflation must end (at-least) locally and there must exist a large ($\sim e^{180} H^{-3}$) volume which thermalizes and ultimately settles to the electroweak vacuum\footnote{We assume no specific mechanism for the end of inflation provided by the inflaton, as this is not necessary for our purposes.}. To describe this post-inflationary era, a finite temperature field theory analysis is required and one must compute tunneling probability of the Higgs towards it's true vacuum by forming a thermal bubble. 

In this paper, apart from pointing out subtleties and assumptions made in this subject in the literature, our primary objective is two-fold: First, to compute the above mentioned statistical quantities for the eternally inflating universe in the presence of the SM Higgs. For this we make use of an integral (kernel) evolution technique~\cite{MM.1} for the distribution of the Higgs (radial degree of freedom in a 4-D Euclidean field space). Second, to describe finite temperature effects after inflation by carefully tracking the rate of bubble formation. Throughout this work, we assume no new physics beyond the minimal SM and work with the (2-loop) renormalized effective Higgs potential with the latest values of relevant parameters~\cite{Particle.data.latest}; we take central values of all parameters, and scan the top mass in the allowed range: $m_{\text{top}} = 172.9 \pm 0.4$ GeV. Also for the sake of simplicity, we assume $H = \text{const.}$ throughout this work; this will be justified because the primary bounds will come at rather low scale inflation values, in which $H$ must be very slowly varying (it requires a very tiny $\varepsilon_{sr}=-\dot{H}/H^2$ to achieve the correct amplitude of density fluctuations).

The organization of the paper is as follows: In section \ref{Steady.state.kernel.non.gaussianities}, we first derive the kernel that propagates the probability distribution for the magnitude (or ``radius") of a multi-component scalar field (relevant to the Higgs doublet) in any Hubble patch. Then to illustrate deviations from Gaussian approximation, we analyze a toy model of 4 fields with a rotationally symmetric ``M" shaped potential and we also simulate a 1+1-dimensional universe. We show comparisons of various statistical quantities obtained from actual simulation, with that from Gaussian approximated and actual steady state distribution obtained using the radial field's kernel. In Section \ref{Higgs.analysis},  in order to have a concrete understanding of the situation with SM Higgs in 3+1 dimensional space-time, we simulate 1+1-dimensional universes with the SM Higgs for different choices of $H$ and central top mass. We provide plots of various statistical quantities, and compare them with ones obtained through steady state distribution using radial field's kernel, and also Gaussian approximated ones. Having obtained confidence in the kernel propagation method, we then use this method to obtain the steady state distribution for our primary subject of interest: the SM Higgs in 3+1 dimensions for various $H$ and different top masses, and use them to get the average size of regions with Higgs within $h_{\text{ht}}$, bounds on inflationary Hubble, and fractal dimension of the eternally inflating universe. 
In section \ref{Temperature.corrections}, we assume the scale of inflation is low enough that the a large universe can be achieved, and then study the post-inflationary era in a simplified way: we account for temperature corrections to the Higgs potential and calculate thermal tunneling probabilities of the Higgs at different temperatures, and show that the SM provides no instabilities even for very high reheating temperatures. In section \ref{Summary} we summarize our results. Finally, in the Appendix we include the relevant SM beta functions.

\section{Evolution of radial field and steady state}\label{Steady.state.kernel.non.gaussianities}

We are interested in the evolution of a spectator Higgs field during inflation. On super-horizon scales it acquires de Sitter fluctuations, with standard deviation (or ``kick") $\kappa$ that acts as a diffusion per Hubble time and is determined by the Hubble scale (we have $\kappa=H/(2\pi)$ in the important case of 3 spatial dimensions, but has different units in other dimensions). We also assume that the field is influenced by its potential $V$. 

To model the evolution of the Higgs field on super-horizon scales, we make use of the usual Fokker-Planck equation. We consider the slow-roll approximation and for now we study general spatial dimensions D. We allow for a multi-component field of length $n$ (the Higgs is a complex doublet with $n=4$) $\vec{\varphi}$, and call the corresponding probability distribution for the field $p(\vec{\varphi},N)$, where $N=Ht$ are the number of e-foldings. The corresponding Fokker-Planck equation for the evolution of the distribution in any Hubble patch is:
\begin{eqnarray}
\dfrac{\partial\,p}{\partial N} = \dfrac{\partial}{\partial\,\vec{\varphi}}\cdot\left[\dfrac{1}{D\,H^2}\dfrac{\partial\,V}{\partial\,\vec{\varphi}}\,p\right] + \dfrac{\kappa^2}{2}\dfrac{\partial^2\,\,p}{\partial\,\vec{\varphi}\,^2}
\label{eq:FokkerPlanck_nfields}
\end{eqnarray}
This holds only within a field domain that allows inflationary energy to dominate and hence inflation to continue. For our purposes we have $V = V(|\vec{\varphi}|) \equiv V(\rho)$, and the inflationary domain is $0 \le \rho \le \rho_{\text{end}}$ where $\rho_{\text{end}}$ is the slow-roll end point. Also, there exists some instability scale $\rho_{\text{ht}}$ in the potential function such that once $\rho$ gets past it, it's classical evolution is to roll down to higher and higher field values which quickly acquire negative energy densities and ultimately around $\rho \simeq \rho_{\text{end}}$ inflation ends. As we shall discuss, a useful region to focus on is in fact $0\leq\rho\leq \rho_{\text{ht}}$, since when we are near the critical Hubble values of interest, the field value $\rho_{\text{ht}}$ is actually quite close to $\rho_{end}$ (see ahead to Fig.~\ref{fig:FastRoll}). 

Solving this PDE requires precise boundary conditions $p(0,N)$ and $p(\rho_{\text{end}},N)$, which can be highly non-trivial~\cite{A.Linde.3,A.Linde.4}, unless the form of potential is highly simplified. Furthermore, it seems impossible to obtain exact solutions for any initial condition $p(\rho,0)$. This can be dealt with by means of numerical integral evolution with support only within the specified domain~\cite{MM.1}. 

\subsection{Kernel for Radial Field}\label{Radial.kernel}

To make progress, it is useful to discretize the time variable and step the evolution through time by use of a kernel. Upon time-discretization, the kernel $K$ that evolves the distribution through one time step $\epsilon$ can be determined by the following chain of reasoning. Firstly, the kernel is implicitly defined by the condition
\begin{equation}
p_{i}\left(|\vec{\varphi}|_{i}\right) = \int \prod^{n}_{a=1}d\varphi^{a}_{i-1}\,K\left(\vec{\varphi}_{i},\vec{\varphi}_{i-1},\epsilon\right)p_{i-1}\left(|\vec{\varphi}|_{i-1}\right),
\end{equation}
The field itself evolves in a stochastic fashion, governed by the Langevin equation
\begin{equation}
\dfrac{\vec{\varphi}_{i} - \vec{\varphi}_{i-1}}{\epsilon} + \dfrac{1}{D H^2}\dfrac{\partial\,V}{\partial\,\vec{\varphi}}\left(|\vec{\varphi}_{i-1}|\right) = \kappa\,\vec{\eta}_{i}
\label{eq:langevindiscrete}
\end{equation}
with $\vec{\eta}_{i}$ being a set of $n$ Normal random variables i.e. having the following probability density
\begin{equation}
\prod^{n}_{a=1}d\eta^{a}_{i}\,P\left(\vec{\eta}_{i}\right) = \left(\prod^{n}_{a=1}d\eta^{a}_{i}\sqrt{\dfrac{\epsilon}{2\pi}}\right)e^{-\frac{\epsilon}{2}\sum_{a}\left(\eta_{i}^{a}\right)^2}.
\end{equation}
Then one can show the appropriate kernel is
\begin{eqnarray}
K\left(\vec{\varphi}_{i},\vec{\varphi}_{i-1},\epsilon\right) &=& \dfrac{1}{\left(2\pi\kappa^2\epsilon\right)^{n/2}}\,\times\nonumber\\
&& e^{-\frac{\epsilon}{2\kappa^2}\sum_{a}\left(\frac{\varphi^{a}_{i}-\varphi^{a}_{i-1}}{\epsilon} + \frac{1}{DH^2}\frac{\partial\,V_{i-1}}{\partial\,\varphi^{a}}\right)^2}
\label{eq:kernel.nfields}
\end{eqnarray}
with support only in $0 \le |\vec{\varphi}|_{i} \le |\vec{\varphi}|_{\text{end}}$, and is trivially obtained by replacing $\eta^{a}_{i}$ for $\varphi^{a}_{i}$. Also, the derivative of potential (or any function of only the radial field $\rho$ for that matter) is
\begin{equation}
\dfrac{\partial\,V}{\partial\,\vec{\varphi}} = \dfrac{\vec{\varphi}}{\rho}\dfrac{d\,V}{d\rho}.
\end{equation}

Since it is overly complicated to track all individual components of the field, we can focus on its radial (or magnitude) $\rho=|\vec{\varphi}|$. This is all we really need as the Higgs potential is symmetric, and it is only the radial value that determines its central dynamics (i.e., if $\rho$ is small then the field is safe, or if $\rho$ is too large then it can be unstable; see ahead to Fig.~\ref{fig:runninglambda} (lower panel) where we replace $\rho\to h$). 

We emphasize that in the case of the SM, even though different components of the Higgs doublet are related by gauge redundancies, one must still use a {\em volume measure} on the 4 component field space, as we explained in Ref.~\cite{Hertzberg:2018kyi}. In order to obtain the kernel for $\rho$ then, we express the field variables in angular coordinates. With $\theta$ and $\alpha_{\mu}$ as the polar angle and the other $n-2$ angles respectively, the integration measure is
\begin{equation}
\prod^{n}_{a=1}d\varphi^{a} = d\rho\,\rho^{n-1}\,d\Omega(\theta,\alpha).
\end{equation}
where $d\Omega$ is the differential solid angle. Throughout this paper, we will absorb the radial measure $\rho^{n-1}$ and total solid angle $\Omega$ into the probability distribution $p$, so that it is normalized as $\int_0^\infty d\rho\,p(\rho)=1$. Now since the whole dynamics must only depend on $\rho$ at any time step, we can always align $\vec{\varphi}_{i}$ and $\vec{\varphi}_{i-1}$ such that $\vec{\varphi}_{i-1}$ points along the polar axis for $\vec{\varphi}_{i}$. With this, we have the following simplification:
\begin{eqnarray}
&&\sum^{n}_{a=1}\left(\dfrac{\varphi^{a}_{i}-\varphi^{a}_{i-1}}{\epsilon} + \dfrac{1}{DH^2}\dfrac{\partial\,V_{i-1}}{\partial\,\varphi^{a}}\right)^2 \nonumber\\
&=& \left(\dfrac{\rho_{i} - \rho_{i-1}}{\epsilon} + \dfrac{1}{DH^2}\dfrac{d\,V_{i-1}}{d\rho}\right)^2\nonumber\\
&-& 2\dfrac{\rho_{i}}{\epsilon}\left(\dfrac{\rho_{i-1}}{\epsilon} - \dfrac{1}{DH^2}\dfrac{d\,V_{i-1}}{d\rho}\right)\left(1 - \cos\theta_{i}\right),
\end{eqnarray}
and we can integrate over the solid angle for any $n$ in general. For the relevant case of 4 fields (Higgs doublet), we have
\begin{equation}
d\Omega = d\theta\,\sin^2\theta\,d\alpha_{1}\,d\alpha_2\sin\alpha_2 \rightarrow 4\pi\,d\theta\,\sin^2\theta
\end{equation}
with $0 \le \theta \le \pi$, and therefore the radial field's kernel for a 4-field system is
\begin{eqnarray}
K\left(\rho_{i},\rho_{i-1},\epsilon\right) &=& \dfrac{e^{-\frac{\epsilon}{2\,\kappa^2}\left(\frac{\rho_{i}}{\epsilon} - z_{i-1}\right)^2}}{\epsilon^2\,\kappa^2}\,\dfrac{I_{1}\left(\frac{\rho_{i}\,z_{i-1}}{\kappa^2}\right)\,e^{-\frac{\rho_{i}\,z_{i-1}}{\kappa^2}}}{\rho_{i}\,z_{i-1}}\nonumber\\
\label{eq:radial.kernel}
\end{eqnarray}
with support only in $0 \le \rho_{i} \le \rho_{\text{end}}$, and
\begin{equation}
z_{i-1} \equiv \dfrac{\rho_{i-1}}{\epsilon} - \dfrac{1}{DH^2}\dfrac{d\,V_{i-1}}{d\rho}.
\end{equation}
Here $I_{1}$ is the modified Bessel function of the $1^{st}$ kind. Therefore we have the following evolution equation for the radial field's distribution
\begin{eqnarray}
p_{i}\left(\rho_{i}\right) &=& \int^{\rho_{\text{end}}}_{0} d\rho_{i-1}\,K\left(\rho_{i},\rho_{i-1},\epsilon\right)\,p_{i-1}\left(\rho_{i-1}\right)\nonumber\\
&=& \prod^{i-1}_{s=0}\int^{\rho_{\text{end}}}_{0}d\rho_{s}\,K\left(\rho_{s+1},\rho_{s},\epsilon\right)\,p\left(\rho_0,0\right).
\end{eqnarray}
This is an iterative matrix multiplication technique (upon discretization of field space) and will converge to the dominant eigenstate of the kernel, which after normalization will give the steady state distribution $\tilde{p}(\rho)$
\begin{equation}
\tilde{p}(\rho) \equiv \dfrac{p(\rho,\infty)}{\int^{\rho_{\text{end}}}_{0}d\rho\,p(\rho,\infty)}.
\label{eq:ssdistrib}
\end{equation}

\subsection{Gaussian Approximation and its Limitations}\label{Mpotential}

A concrete way to analyze the Fokker-Planck equation is to study its moments. For example, the equation for the variance $\sigma^2$ can be obtained by multiplying the Fokker-Planck eq.~\eqref{eq:FokkerPlanck_nfields} with $\vec{\phi}\,^2$ and integrating over the fields $\vec{\phi}$:
\begin{equation}
\dfrac{d}{dN}\sigma^2 + \dfrac{2}{n\,D\,H^2}\langle\vec{\phi}\cdot\nabla_{\phi}V\rangle = \kappa^2.
\label{eq:sigma_gaussian}
\end{equation}
Similarly one can obtain equations of motion for higher moments. 

To simplify things however, very often in the literature a Gaussian approximation is made:
\begin{equation}
p^{(0)}(\rho,N) = \dfrac{\rho^{n-1}}{\sqrt{(2\pi)^n}\sigma^n(N)}\exp\left[-\dfrac{\rho^2}{2\,\sigma^2(N)}\right].
\label{eq:Gaussian.approximation}
\end{equation}
where, as mentioned earlier, we have absorbed the radial integration measure $\rho^{n-1}$ in the distribution. The only quantity then, is the variance $\sigma^2$ which obeys the above simple ODE.

An important point to note however, is that for potentials like that of the SM Higgs and especially under a Gaussian ansatz, there exists a critical Hubble $H_{\text{cr}}^{(0)}$ (that depends on the parameters of the potential), such that although for $H < H_{\text{cr}}^{(0)}$ this Gaussian approximated distribution $p^{(0)}$ tends to a constant function even if one had no a-priori knowledge of the steady state-ness in eternal inflation; for $H > H_{\text{cr}}^{(0)}$, $p^{(0)}$ grows indefinitely. In any case, people often simplify the analysis by evolving it for some finite e-foldings (typically $\sim 60$ in the case of SM Higgs for our universe) and use this to calculate various quantities, e.g. probability of obtaining large volumes with field within the domain $0 \le \rho \le \rho_{\text{ht}}$ etc.~\cite{Espinosa.Riotto.1,Zurek.1,Zurek.2,Kohri.1,Kohri.2}. However in either case, the simplification made is significant. First of all it is obvious that due to a finite $\rho_{\text{end}}$, for any $H$ there exists a steady state distribution $\tilde{p}(\rho)$ which cannot have support outside of $\rho > \rho_{\text{end}}$. Then, to leading order the Gaussian approximation can be valid \textit{in principle}, only as long as $\sigma(N) \lesssim \rho_{\text{end}}$. For the latter case $H > H_{\text{cr}}^{(0)}$, this eventually breaks completely and therefore the approximation is not meaningful for all $N$. For the former case $H < H_{\text{cr}}^{(0)}$, even though it could be that $\sigma(\infty) < \rho_{\text{end}}$, the non-Gaussianity in the actual steady state distribution could be significant enough for there to be corrections in the estimated quantities. To demonstrate this, we consider the following ``M" shaped potential:
\begin{equation}
V(\rho) = \dfrac{m^2}{2}\rho^2 - \dfrac{\lambda}{4}\rho^4 + \text{constant}.
\end{equation}
(ironically, this is upside down from the classical Higgs potential, but it qualitatively captures the RG corrected Higgs potential since it in fact rises, then goes negative at large values, just like this). This leads to the following equation for the variance in the Gaussian approximation
\begin{equation}
\dfrac{d}{dN}\sigma^2 + \dfrac{2}{DH^2}\left[-\lambda(n+2)\sigma^4 + \sigma^2m^2\right] = \kappa^2
\label{eq:toysigma_evolve}
\end{equation}
which can be re-written as
\begin{equation}
\dfrac{d}{dN}\sigma^2 - \dfrac{2\,\lambda(n+2)}{DH^2}\left(\sigma^2 - \sigma^2_{+}\right)\left(\sigma^2 - \sigma^2_{-}\right) = 0.
\label{eq:toysigma_evolvebetter}
\end{equation}
For the sake of comparison with actual 1-D simulations, we work in 1 spatial dimension (although the argument holds for any spatial dimension in general) for which we have $\kappa = 1/\sqrt{2\pi}$~\cite{MM.1}. This gives
\begin{eqnarray}
\sigma^2_{\pm} &=& \dfrac{m^2}{2\,\lambda(n+2)}\left[1 \pm \sqrt{1 - \left(\dfrac{H}{H_{\text{cr}}^{(0)}}\right)^2}\right],\nonumber\\
H^{(0)}_{\text{cr}} &=& \left(\dfrac{\pi\,m^4}{\lambda(n+2)}\right)^{\frac{1}{2}};
\end{eqnarray}
with $H^{(0)}_{\text{cr}}$ the above mentioned critical Hubble. The solution is trivially obtained (with $\sigma(0) \approx 0$):
\begin{equation}
\sigma^2(N) = \dfrac{\sigma^2_{+}\sigma^2_{-}\left(1 - e^{-\gamma N}\right)}{\sigma^2_{+} - \sigma^2_{-}e^{-\gamma N}}
\label{eq:toysigma_soln3}
\end{equation}
where 
\begin{equation}
\gamma \equiv \frac{2\,\lambda(n+2)}{H^2}(\sigma^2_{+} - \sigma^2_{-}).
\end{equation}
Now for $H \le H^{(0)}_{\text{cr}}$, we have an attractive fixed point $\sigma_{-}$ and thus $\sigma(\infty) = \sigma_{-}$ dictating a constant distribution\footnote{This is true for any $\sigma(0) < \sigma_{+}$}. In this case, the Gaussian approximation is valid in principle since we have $\sigma_{-} < \rho_{\text{end}}$ where $\rho_{\text{end}}$ marks the slow-roll end point
\begin{eqnarray}
|V''(\rho_{\text{end}})| &\simeq& 9\,H^2\nonumber\\
\implies \rho_{\text{end}} &\simeq& \sqrt{\dfrac{9\,H^2 + m^2}{3\,\lambda}},
\label{eq:rhoend}
\end{eqnarray}
and neglecting the effect of the finite-ness of $\rho_{\text{end}}$ is warranted, at least in principle, to first order in approximations\footnote{The factor of $9$ here is not of much significance. Our results are very insensitive to increments in the slow-roll end point.}. However on the other hand for $H > H^{(0)}_{\text{cr}}$, there is no real fixed point and the distribution grows indefinitely $\sigma(\infty) = \infty$. The Gaussian approximation therefore completely breaks. In reality a finite $\rho_{\text{\text{end}}}$ would lead to a steady state distribution $\tilde{\rho}(\rho)$ which we can easily obtain numerically using the radial field's kernel derived earlier.

\subsection{Comparisons}

We now present some numerical results for various important quantities. For a 1+1-dimensional simulation, we have used our network-I as we outlined in detail in Ref.~\cite{MM.1}. We begin with one cell with 4 field values all at zero and double the number of cells at each step. The field values in the daughter cells are dictated by Langevin equation \eqref{eq:langevindiscrete} with $\epsilon = \ln 2$, and are produced only if the radial field $\rho$ in the parent cell was less than the slow-roll end value $\rho_{\text{end}}$. For the kernel evolution, we begin with a delta distribution $p(\rho,0) = \delta(0)$ and evolve it until the normalized distributions at the final step and one previous step (with $\epsilon = \ln 2$) differed from each other by only 1 part in $10^3$. In order to converge to steady state in a few e-foldings in our simulations for this case of M potential, we carefully choose our parameters:

For $H \ll H^{(0)}_{\text{cr}}$, we can use Gaussian approximation to estimate the scaling of the time of convergence. From eq.~\eqref{eq:toysigma_soln3}, the number of e-foldings to converge to $\sigma_{-}$ goes like
\begin{equation}
N_{*} \sim \dfrac{1}{\gamma} \sim \dfrac{H^2}{m^2} = \dfrac{H^2}{\lambda\,\rho_{\text{ht}}}.
\end{equation}
On the other hand for $H \gg H^{(0)}_{\text{cr}}$, we can use classical equation of motion to estimate the time it takes for the field to roll to $\rho_{\text{end}}$. A straightforward calculation gives the following
\begin{equation}
N_{*} \sim \dfrac{H}{\kappa\sqrt{\lambda}}.
\end{equation}
We model different scenarios keeping $\lambda$ fixed at $0.06\,H^2$ while varying $\rho_{\text{ht}}$ (in units of $\kappa$) such that we always have $N_{*} \lesssim \mathcal{O}(10)$ . For any given scenario, our results are very insensitive to increments in the slow-roll end values. c.f. eq.~\eqref{eq:rhoend}. 
Note that the Gaussian approximation is so simple that we carry it out in the continuum limit, while the full simulations and kernel evolution are more complicated, so they have a discrete time step, as mentioned above.

\begin{figure}[t]
\centering
\includegraphics[width=1\columnwidth]{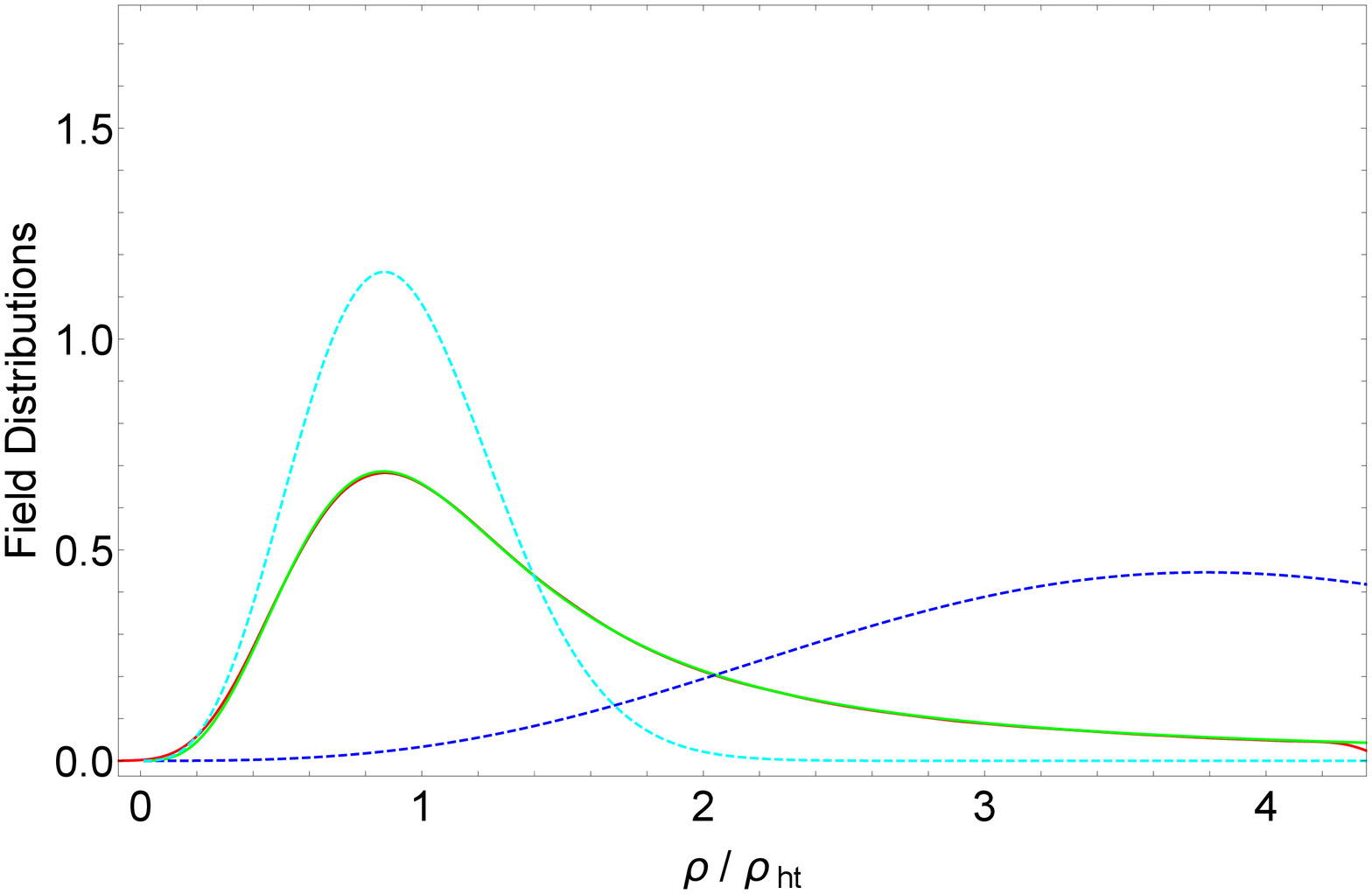}\vspace{0.5cm}\\
\includegraphics[width=1\columnwidth]{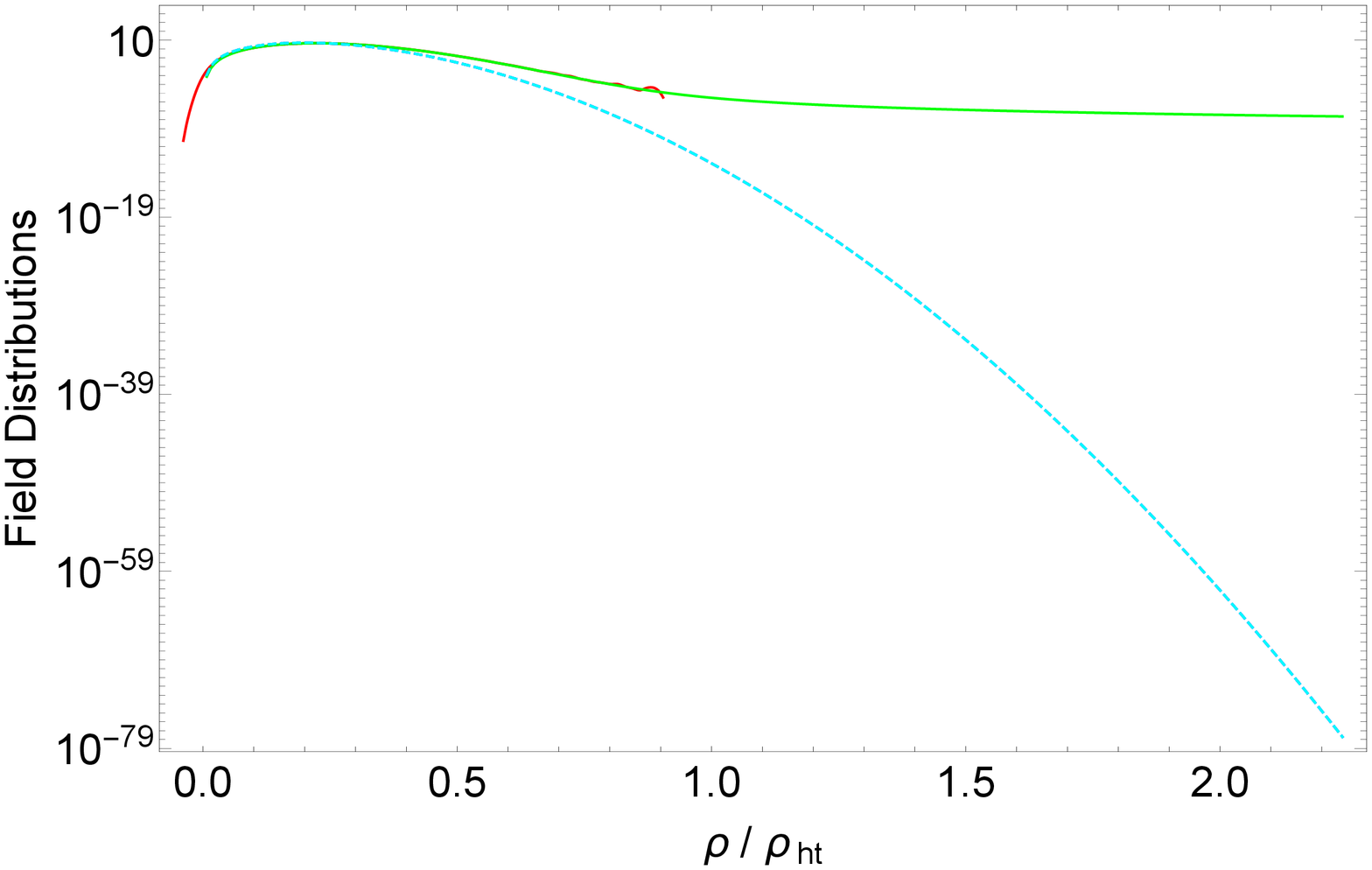}
\caption{Upper panel: Field distribution in a Hubble patch for $H \approx 2.16\,H^{(0)}_{\text{cr}}$. Red curve is the actual simulation result after $18$ 2-foldings, dashed green curve is the steady state distribution obtained from kernel evolution, and dashed cyan, blue curves are Gaussian approximations with their widths truncated at $\rho_{\text{ht}}/2$ and $\rho_{\text{end}}/2$ respectively, and $\rho_{\text{end}} \approx 2.18\,\rho_{\text{ht}}$.
Lower panel: Field distribution in a Hubble patch for $H \approx 0.54\,H^{(0)}_{\text{cr}}$. Color coding is the same as figure~\ref{fig:Mpotentialfielddistribution1}. For the Gaussian approximated distribution, $\sigma(\infty) = \sigma_{-} \approx 0.11\,\rho_{\text{ht}}$. This is all in 1+1-dimensions}
\label{fig:Mpotentialfielddistribution1} 
\end{figure}

To begin, we first show plots of fied distributions for the two cases of $H$. Fig.~\ref{fig:Mpotentialfielddistribution1} (upper panel) compares the field distribution obtained from simulation after $18$ 2-foldings (red curve) with the actual steady state distribution obtained from kernel evolution (green curve), and Gaussian approximations (cyan, blue) for an $H > H^{(0)}_{\text{cr}}$. In order to force a finite bounded distribution for the latter (otherwise spreading indefinitely), we show two illustrations of truncating the standard deviation of the radial field at the hilltop and the slow-roll end point, i.e. $\sigma(\infty) \equiv \rho_{\text{ht}}/2$ and $\sigma(\infty) \equiv \rho_{\text{end}}/2$ respectively.

Fig.~\ref{fig:Mpotentialfielddistribution1} (lower panel), on the other hand, compares simulation result against the steady state and Gaussian approximated ($\sigma(\infty) = \sigma_{-}$) distributions for an $H < H^{(0)}_{\text{cr}}$. Here even though a Gaussian approximation is allowed, deviations from it in the actual distribution are large. In either case, our steady state distribution is in excellent agreement with the actual simulation result.

\begin{figure}[t]
\centering
\includegraphics[width=1\columnwidth]{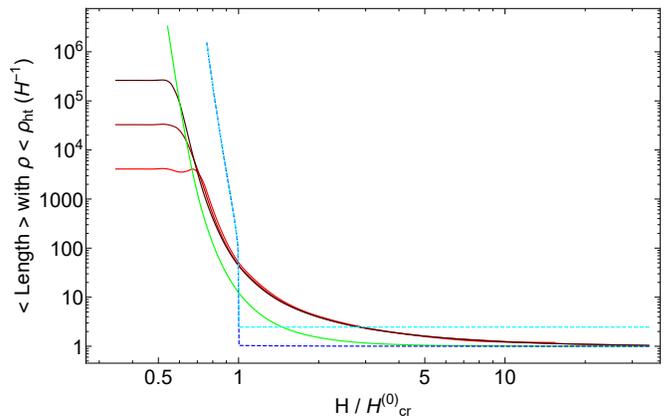}
\caption{\label{fig:Mpotentialaveragevolume} Average length of chains with radial field within the hilltop in 1+1-dimensions. Red curves are from simulations at $12$, $15$, and $18$  $2$-foldings in increasing order of darkness, while the green and cyan/blue are from actual steady state and Gaussian approximated distributions respectively with the assumption of completely uncorrelated Hubble patches \eqref{eq:average.length.uncorrelated}.}
\end{figure}

Now, one of the interesting and necessary quantities that we must compute, is the average size of regions that contain the radial field $\rho$ within the hilltop value $\rho_{\text{ht}} = m/\sqrt{\lambda}$. As can be seen from above distribution plots, the non-Gaussianity in the actual (steady state) distribution is important to track. Fig.~\ref{fig:Mpotentialaveragevolume} shows the average length of inflating chains in 1+1-dimensions, computed from steady state distribution obtained from kernel evolution (green curve), and Gaussian approximated distributions (cyan and blue dashed curves), as compared with the actual simulation result (red curves). The former two are obtained assuming completely uncorrelated patches, giving the following
\begin{eqnarray}
\langle\,\text{Length}\,\rangle &=& \dfrac{1}{1-f}\,H^{-1};\nonumber\\
f &=& \int^{\rho_{\text{ht}}}_{0} d\rho\;\tilde{p}(\rho)
\label{eq:average.length.uncorrelated}
\end{eqnarray}
Here again for $H > H^{(0)}_{\text{cr}}$, we have the above two considerations of truncating the width of Gaussian approximated distribution at $\rho_{\text{ht}}/2$ and $\rho_{\text{end}}/2$. It is apparent from the figure that not only is the Gaussian approximated result further from the steady state result, the latter is much closer to the actual simulation result especially for smaller $H$, and gets even closer for larger lengths.

\begin{figure}[t]
\centering
\includegraphics[width=1\columnwidth]{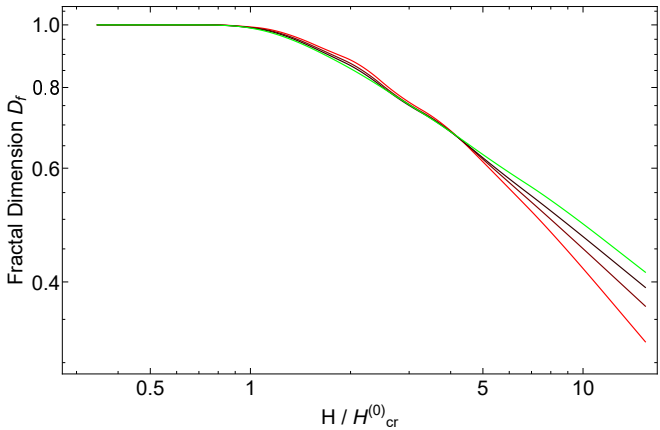}
\caption{\label{fig:Mpotentialfractaldimension} Fractal dimension of 1+1-dimensional universe with $\rho < \rho_{\text{ht}}$. Red curves are obtained from 1-D simulation at $12$, $15$, and $18$ 2-foldings respectively while green curve is obtained from kernel evolution at steady state. c.f. eq.~\eqref{eq:greenvolume}.}
\end{figure}

Another interesting quantity to consider is the fractal dimension of chains that are within the instability scale i.e. $\rho < \rho_{\text{ht}}$, defined as
\begin{equation}
D_{f} = \lim_{N \rightarrow \infty}\dfrac{\ln\,L_{< \rho_{\text{ht}}}(N)}{N}
\label{eq:fractal.dimension}
\end{equation}
where $L_{< \rho_{\text{ht}}}(N)$ is the length of inflating regions with field values within the instability point. With time discretization $\epsilon$ and total number of steps $s$ such that $N = \epsilon\,s$, it is equal to the following
\begin{eqnarray}
L_{< \rho_{\text{ht}}}(N) = e^{\epsilon\,s} \prod^{s-1}_{i=0}\int^{\rho_{\text{ht}}}_{0}d\rho_{i}\,K\left(\rho_{i+1},\rho_{i},\epsilon\right)\,p\left(\rho_0,0\right).\nonumber\\
\label{eq:greenvolume}
\end{eqnarray}
Fig.~\ref{fig:Mpotentialfractaldimension} compares the fractal dimension with $\rho < \rho_{\text{ht}}$ obtained from 1+1-dimensional simulations after $12$, $15$, and $18$ $2$-foldings, respectively, with that from kernel propagation at steady state. The convergence of the simulation towards steady state result obtained from kernel evolution is apparent from this plot.

Having laid out a quantitative analysis for the M potential, all the while elucidating some subtleties and simplifying assumptions made in the literature, and also establishing validity of radial kernel~\eqref{eq:radial.kernel} to compute various quantities of interest, we now analyze the eternal inflation scenario with the SM Higgs.

\section{Analysis with the Standard Model Higgs}\label{Higgs.analysis}

\begin{figure}[t]
\centering
\includegraphics[width=1\columnwidth]{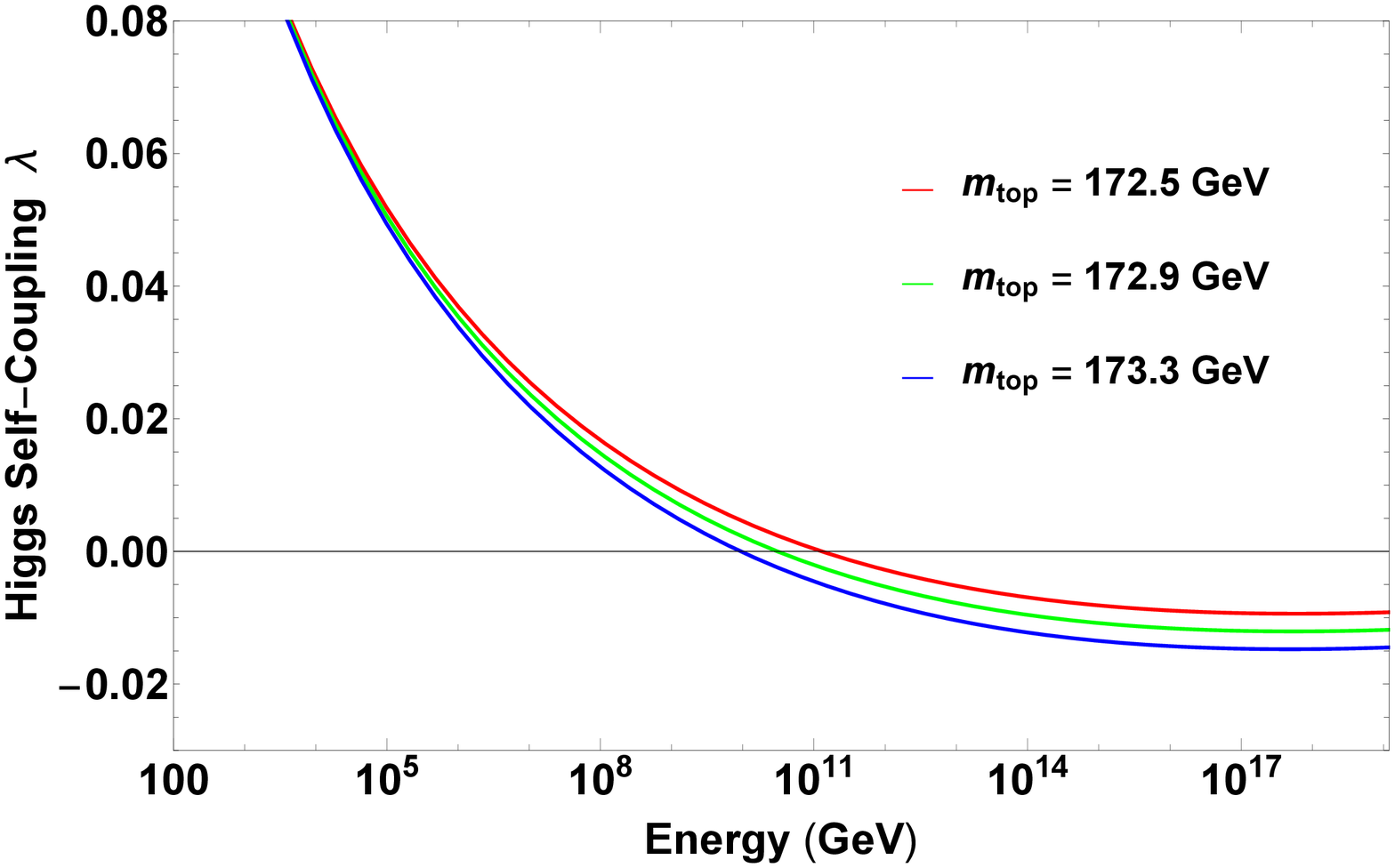}\vspace{0.5cm}\\
\includegraphics[width=1\columnwidth]{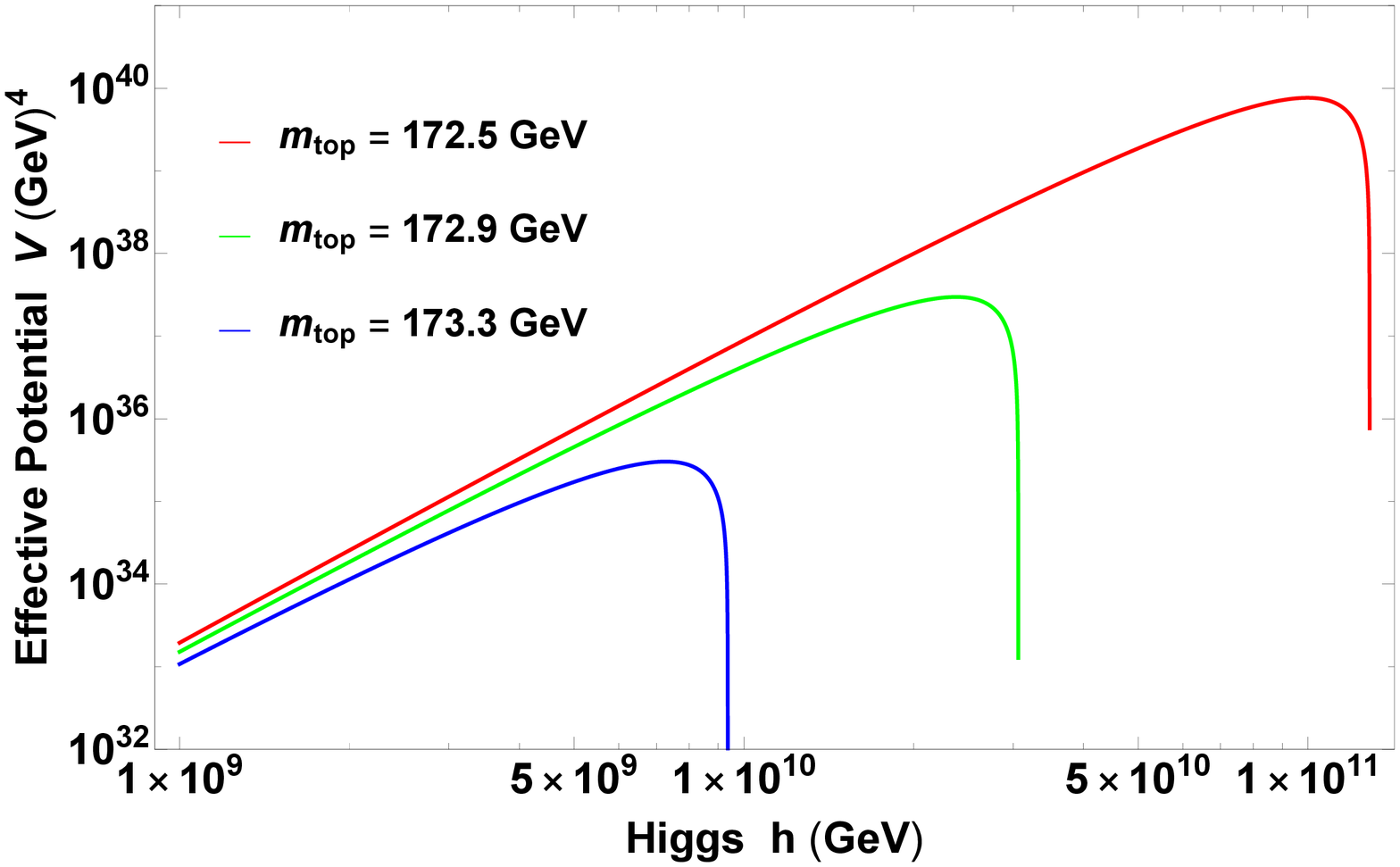}
\caption{\label{fig:runninglambda} Upper panel: The running of self coupling $\lambda$ as a function of energy within the minimal SM for 3 different values of the top mass, which spans the currently favored top mass window. 
Lower panel: The corresponding effective Higgs potential $V$ as a function of Higgs value $h$ on a log-log scale.}
\end{figure}

The SM potential for the Higgs $V=-\mu^2 h^2+\lambda\, h^4/4$ is known to pick up corrections from loops. These lead to logarithmically slow changes in $\lambda$ as we go to high energies. For $m_{\text{top}} = 172.9 \pm 0.4$ GeV, the Higgs self-coupling $\lambda$ becomes negative at around $\sim 10^{10}-10^{11}$ GeV within the minimal SM. Fig.~\ref{fig:runninglambda} (upper panel) shows the RG evolution of $\lambda$ for the central, upper, and lower values of top mass, at two loop order. For all other relevant parameters, we have taken their central values~\cite{Particle.data.latest}: $m_h=125.1$\,GeV, $\alpha_{em}=1/127.9$, $\alpha_s=0.1181$, $\sin^2(\theta_w)=0.23122$ (all evaluated at Z mass). The 2-loop beta functions for $\lambda$, the top quark Yukawa coupling, gauge couplings, and the anomalous dimension are provided in Appendix ~\ref{appA}.

By replacing the energy scale by $E\to h$, the corresponding renormalized effective Higgs potential can be written as
\begin{equation}
V(h) = \dfrac{\lambda(h)}{4}\,G(h)^4\,h^4,
\label{eq:higgseffectivepotential}
\end{equation}
(with $G$ related to wave-function renormalization). The potential is plotted in Fig.~\ref{fig:runninglambda} (lower panel) on a log-log scale. The point at which this reaches a maximum $V'(h)=0$, we call the ``hill top" value $h_{\text{ht}}$. It is found to be given by
\begin{eqnarray}
h_{\text{ht}} =
\left\{
	\begin{array}{ll}
		1.0 \times 10^{11} \; \text{GeV} &  \mbox{for} \;\; m_{\text{top}} = 172.5 \; \text{GeV} \\
		2.4 \times 10^{10} \; \text{GeV}  &  \mbox{for} \;\; m_{\text{top}} = 172.9 \; \text{GeV} \\
		7.3 \times 10^{9} \; \text{GeV}  &  \mbox{for} \;\; m_{\text{top}} = 173.3 \; \text{GeV} \\
	\end{array}
\right.\,\,\,
\end{eqnarray}

There exist ideas in the literature as to how to ``cure" the potential by adding new degrees of freedom that alter the Higgs beta functions (such as \cite{EliasMiro:2012ay,Hertzberg:2012zc}), however we will not pursue that here. Our focus is to put the minimal SM on trial to ever increasing energies and see what constraints on inflation, etc, follow.

\subsection{Gaussian Approximation and its Limitations}

For Gaussian approximation, we can proceed similarly as before. Under Gaussian ansatz~\eqref{eq:Gaussian.approximation} and the renormalized Higgs potential, we have the following simple differential equation for the variance
\begin{equation}
\dfrac{d}{dN}\sigma^2 + \dfrac{1}{2\,D\,H^2}\langle\,h\,\dfrac{dV}{dh}\rangle\,(\sigma^2) = \dfrac{H^2}{4\,\pi^2}.
\label{eq:sigma_gaussian2}
\end{equation}
For all cases when the variance asymptotes to a constant eventually, i.e. $H \le H^{(0))}_{\text{cr}} $, we may set $d\sigma^2/dN$ equal to zero giving
\begin{equation}
H^4 = \dfrac{2\,\pi^2}{D}\langle\,h\,\dfrac{dV}{dh}\rangle\,(\sigma^2).
\end{equation}
Thus, we have
\begin{equation}
H^{(0)}_{\text{cr}} = \left(\dfrac{2\,\pi^2}{D}\,\left.\langle\,h\,\dfrac{dV}{dh}\rangle\right|_{\text{max}}\right)^{1/4}.
\end{equation}
and for all other Hubble values smaller than this critical value, we have two roots (fixed points) as before with the smaller being an attractor. Numerically, we obtain the following values of critical Hubble for the three values of top mass and $D=3$:
\begin{eqnarray}
H^{(0)}_{\text{cr}} =
\left\{
	\begin{array}{ll}
		1.4 \times 10^{10} \; \text{GeV} &  \mbox{for} \;\; m_{\text{top}} = 172.5 \; \text{GeV} \\
		3.5 \times 10^{9} \; \text{GeV}  &  \mbox{for} \;\; m_{\text{top}} = 172.9 \; \text{GeV} \\
		1.1 \times 10^{9} \; \text{GeV}  &  \mbox{for} \;\; m_{\text{top}} = 173.3 \; \text{GeV} \\
	\end{array}
\right.\,\,\,\,\,
\end{eqnarray}
We note that this says for $H>H^{(0)}_{\text{cr}}$ there is no fixed point. 
However, as we will again see, when studied more precisely beyond the Gaussian approximation, a steady state is realized for any $H$. However, the sizes of habitable patches will be strongly dependent on the value of $H$. On the other hand, for $H<H^{(0)}_{\text{cr}}$, the Gaussian approximation does give a steady state, though in the literature the distribution is often just cut off at $N\approx60$ instead.

\subsection{1+1 Dimensional Simulation with Higgs and Comparisons}

Now let us move beyond the overly simple Gaussian approximation towards a full treatment of the problem. In order to validate our analysis concretely, we begin by simulating (using our network-I as mentioned earlier) 1+1-dimensional inflating universes with the SM Higgs, for the central value of top mass and different $H$. With random walk step $\kappa$, we scale the field(s) and the potential as
\begin{eqnarray}
\vec{z} &\equiv& \dfrac{\vec{h}}{h_{\text{ht}}}\nonumber\\
W &\equiv& \dfrac{ V}{H^2\,h_{\text{ht}}^2}
\end{eqnarray}
giving the following (discretized-) Langevin equation for the fields' evolution in any Hubble patch
\begin{equation}
\dfrac{\vec{z}_{i} - \vec{z}_{i-1}}{\epsilon} + \dfrac{\vec{z}_{i-1}}{z_{i-1}}\dfrac{\partial\,W}{\partial\,z}\left(z_{i-1}\right) = \dfrac{\kappa}{h_{\text{ht}}}\vec{\eta}_{i},
\label{eq:langevindHiggs1D}
\end{equation}
and the slow-roll end point is obtained as before
\begin{equation}
W''(z_{\text{end}}) \simeq 9.
\end{equation}
To mimick some of the aspects of the desired 3+1-dimensional scenario, we simulate universes with different numerical values of $H$ ranging from $0.1\,h_{\text{ht}}$ to $10\,h_{\text{ht}}$, and simultaneously $\kappa$ ranging from $0.1\,h_{\text{ht}}/2\pi$ to $10\,h_{\text{ht}}/2\pi$. Note that with these choices of parameters, it is practically impossible to achieve steady state in simulations since we would require a large number of 2-foldings. Therefore we obtain results from kernel propagation both for $18$ 2-foldings (which is the maximum amount we go in our simulations), and steady state. To obtain steady state, we propagate the distribution with the appropriate kernel until they start to deviate from each other only by 1 part in $10^3$, as mentioned before. Also for Gaussian approximation and $D=1$, we have $H^{(0)}_{\text{cr}} = 4.54 \times 10^9$ GeV for the central value of top mass. We again note that the simulation and kernel methods have a discrete time step of $\epsilon=\ln 2$, while the Gaussian approximation is in the continuum limit.

\begin{figure}[t!]
\centering
\includegraphics[width=1\columnwidth]{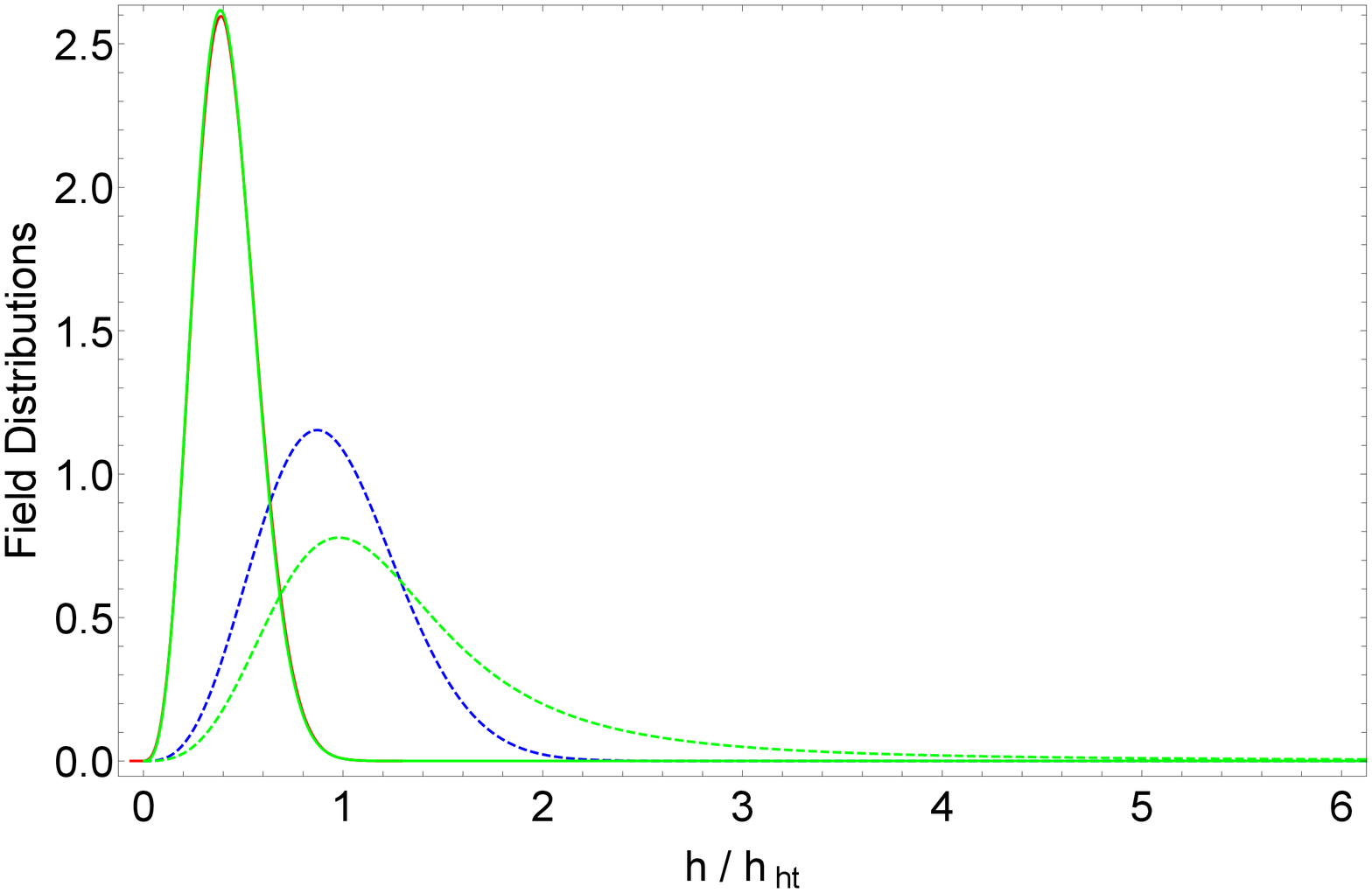}\vspace{0.5cm}\\
\includegraphics[width=1\columnwidth]{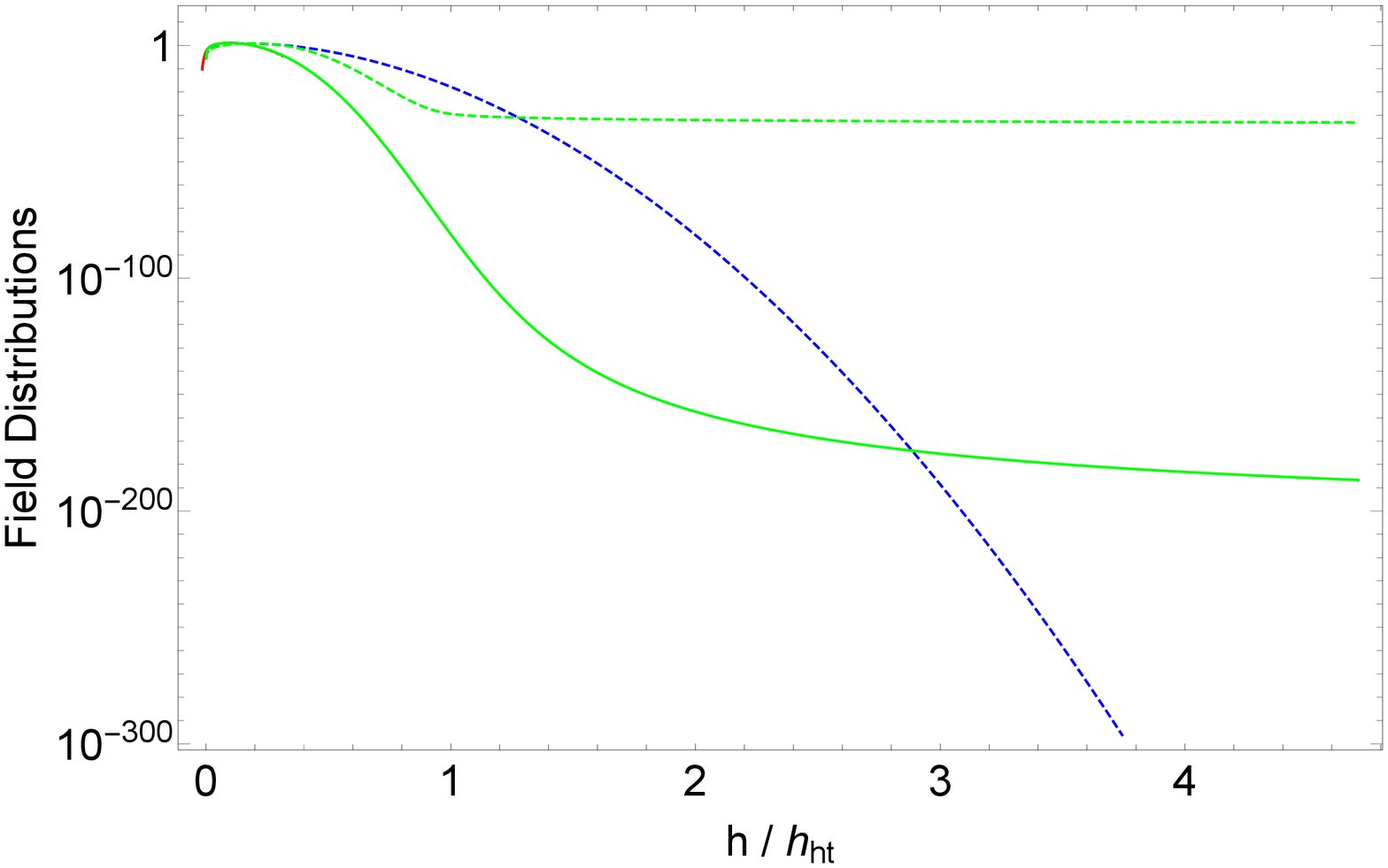}
\caption{Upper panel: Field distribution in a Hubble patch for $H = 2.09\,H^{(0)}_{\text{cr}}$ (i.e., $H > H^{(0)}_{\text{cr}}$) in 1+1-dimensions. Solid red and green curves are from actual simulation and kernel propagation after $18$ 2-foldings respectively, dashed green curve is the steady state distribution obtained from kernel propagation, while dashed blue curve is the Gaussian approximation with the variance truncated at $60$ e-foldings: $\sigma(60) \approx 1.2 \times 10^{10}$ GeV. 
Lower panel: Field distribution in a Hubble patch for $H \approx 0.52\,H^{(0)}_{\text{cr}}$ (i.e., $H<H^{(0)}_{\text{cr}}$) in 1+1-dimensions. Solid red and green curves are from actual simulation and kernel propagation after $18$ 2-foldings respectively, dashed green curve is the steady state distribution obtained from kernel propagation, while dashed blue curve is the Gaussian approximation with $\sigma(\infty) \approx 2.39 \times 10^9$ GeV.}
\label{fig:Higgs1Dfielddistribution1} 
\end{figure}

First, we present field distributions for both cases $H > H^{(0)}_{\text{cr}}$ in Fig.~\ref{fig:Higgs1Dfielddistribution1} (upper panel) and $H < H^{(0)}_{\text{cr}}$ in Fig.~\ref{fig:Higgs1Dfielddistribution1} (lower panel). The excellent agreement between actual simulation result and kernel propagation is evident from the red and dashed green curves (after $18$ 2-foldings). For the former case of $H > H^{(0)}_{\text{cr}}$ when Gaussian approximated $\sigma(N)$ blows up eventually, we evolve it until $N = 60$ as is usually done in the literature. On the other hand for the latter case of $H < H^{(0)}_{\text{cr}}$, we show the field distribution on a log-linear scale to make deviations from Gaussian approximation apparant.

\begin{figure}[t!]
\centering
\includegraphics[width=1\columnwidth]{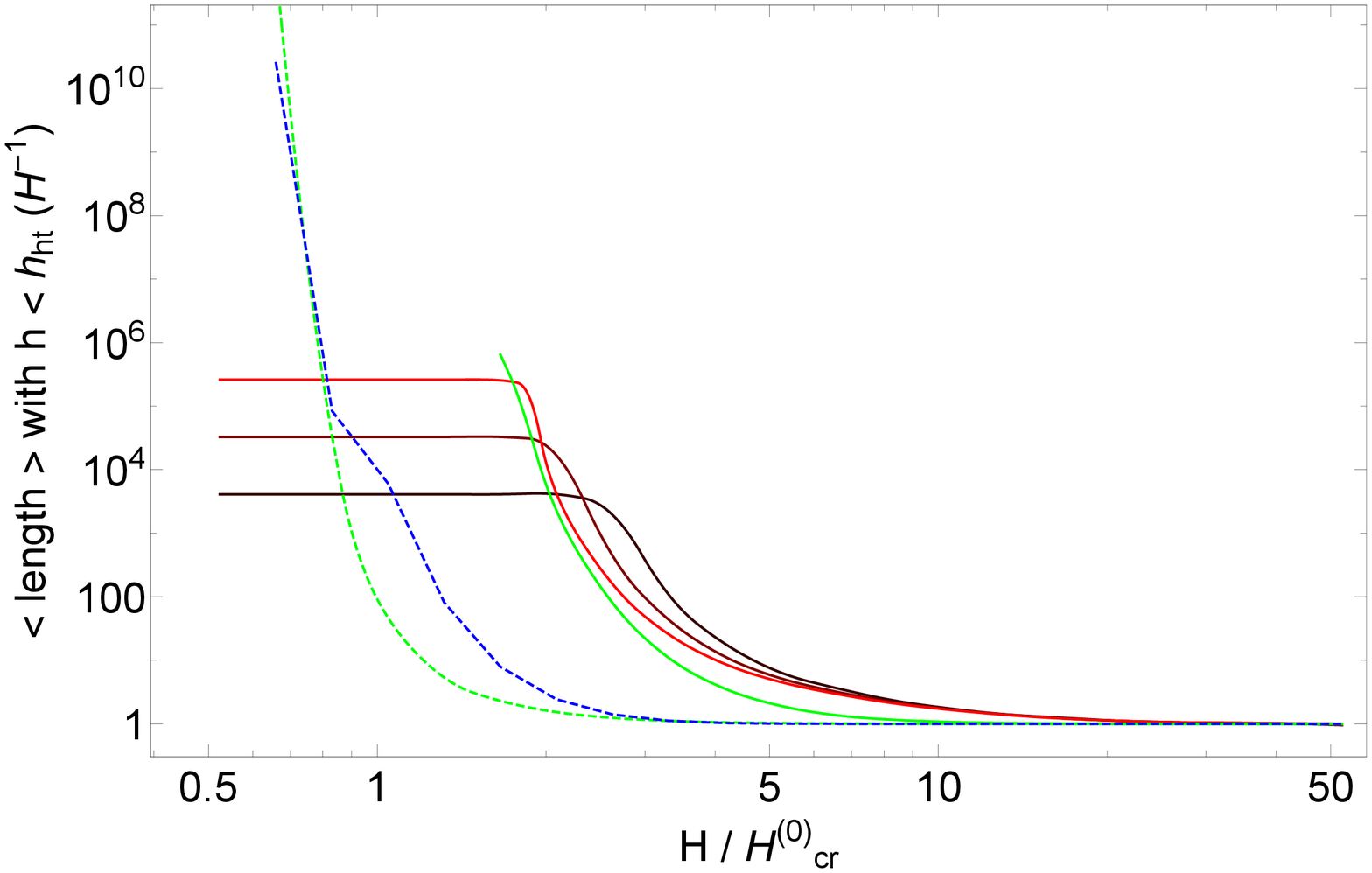}\vspace{0.5cm}\\
\includegraphics[width=1\columnwidth]{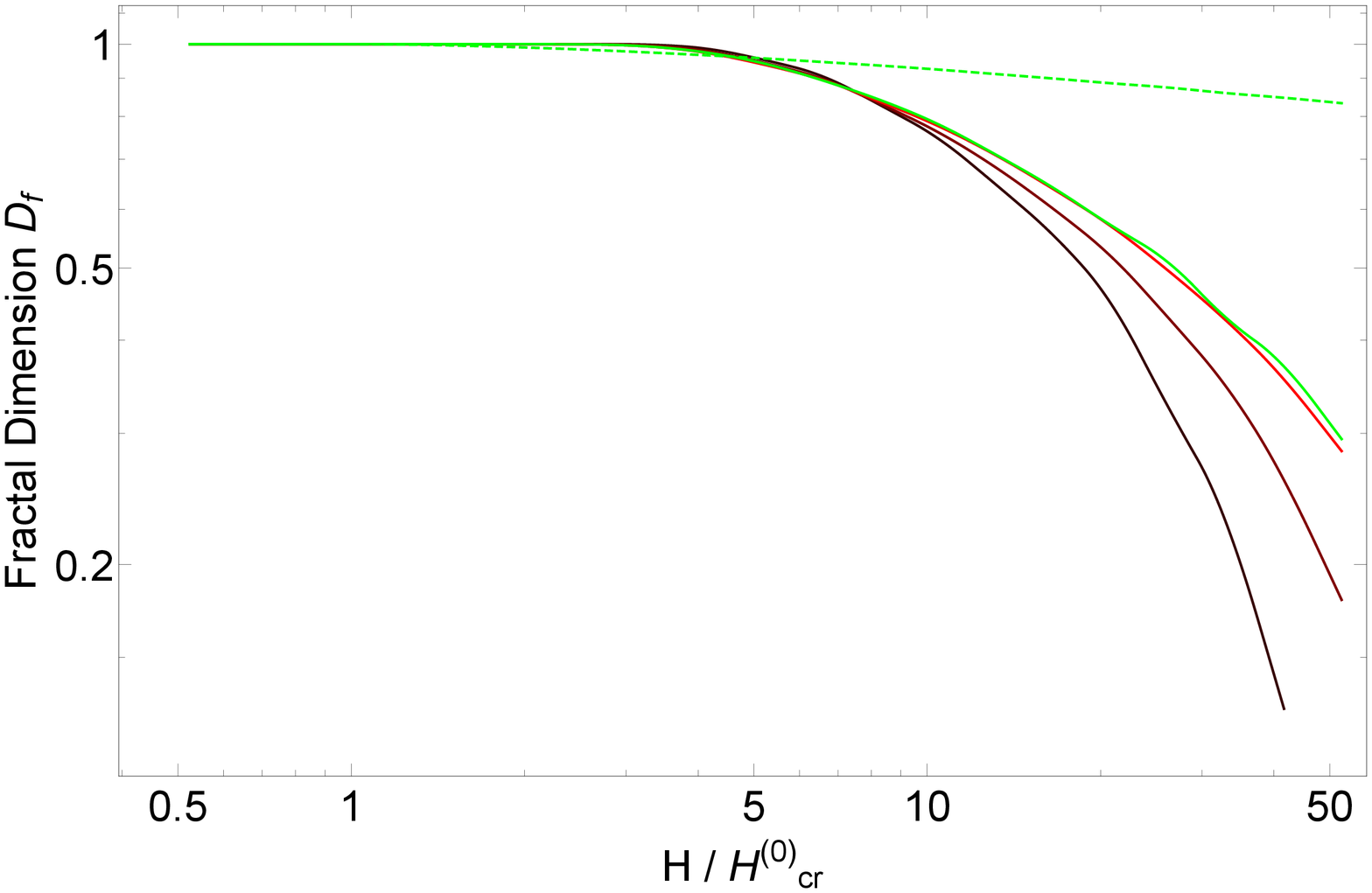}
\caption{\label{fig:Higgs1Daveragevolume} Upper panel: Average length of chains with Higgs within the hilltop in 1+1-dimensions. As before, solid red curves are from actual simulation; while green and dashed blue are from kernel propagation (solid green corresponding to $18$ 2-foldings and dashed green corresponding to steady state) and Gaussian approximated distribution with $\sigma$ truncated at $\sigma(60)$ for $H > H^{(0)}_{\text{cr}}$ respectively, under the assumption of uncorrelated patches.
Lower panel: Fractal dimension of regions containing Higgs within the hilltop. Solid red curves are from actual simulations after $12$, $15$, and $18$ 2-foldings (in increasing brightness). Solid green is from kernel evolution after $18$ 2-foldings, while dashed green is at steady state.}
\end{figure}

The average length of regions containing Higgs within it's hilltop value, is shown in Fig.~\ref{fig:Higgs1Daveragevolume} (upper panel). As before, under the assumption of uncorrelated patches, average length obtained using steady state distribution is given as before (c.f. eq. ~\eqref{eq:average.length.uncorrelated}) with the appropriate kernel for the Higgs potential.
Note that even though the actual average size (red curves) is bigger in reality (due to correlations between nearby Hubble patches), the assumption of uncorrelated patches (solid green curve) gets closer to the former especially for very large sizes. Finally, the fractal dimension of these regions containing Higgs within the hilltop given as before (c.f. eq. ~\eqref{eq:greenvolume} and~\eqref{eq:fractal.dimension}), is shown in Fig.~\ref{fig:Higgs1Daveragevolume} (lower panel).

\subsection{Full Analysis in 3+1 Dimensions}

\begin{figure}[t!]
\centering
\includegraphics[width=1\columnwidth]{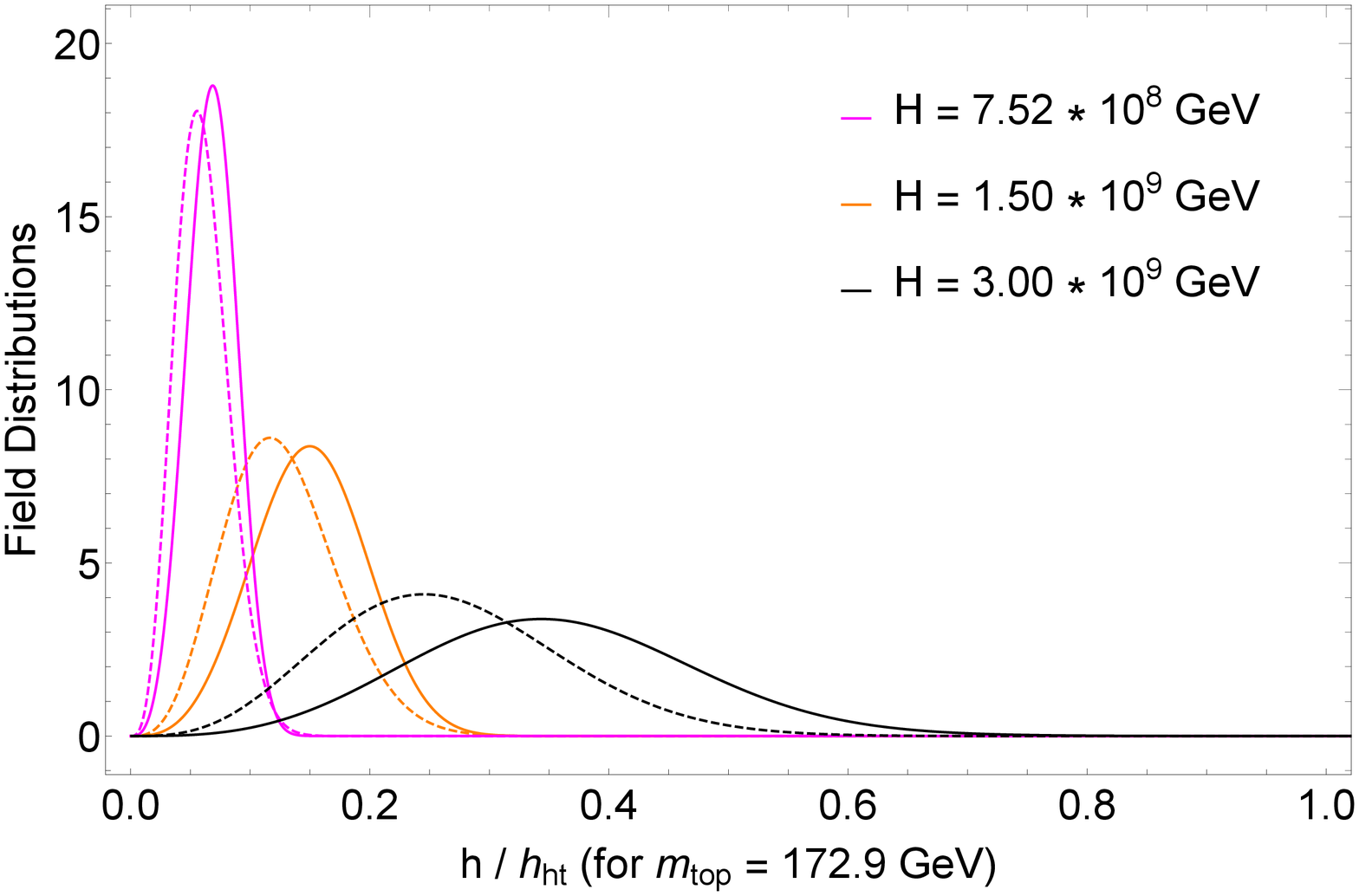}\vspace{0.5cm}\\
\includegraphics[width=1\columnwidth]{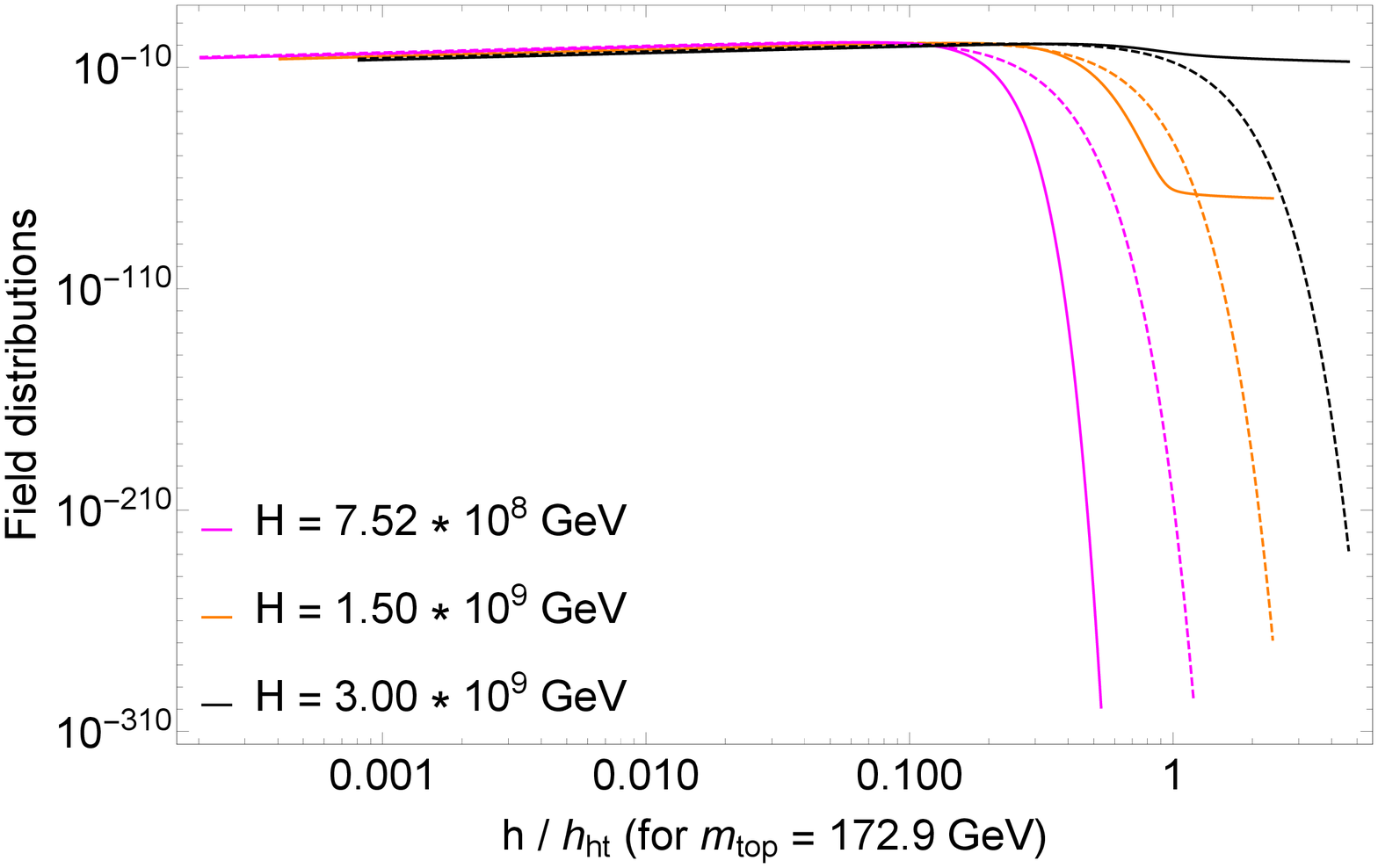}
\caption{\label{fig:higgsdistributions} Upper panel: Higgs dynamics in 1+1-dimensions. Solid curves are some of the steady state distributions for different Hubble values close to the one required for observable universe, while the corresponding dashed curves are the Gaussian approximated distributions with $\sigma$ truncated at $N = 60$ for $H > H^{(0)}_{\text{cr}}$. Lower panel: Same curves, but on a log-log scale to emphasize the tails.}
\end{figure}

Having shown the validity of our kernel evolution technique concretely, we now move on the important case of $3+1$-dimensions. Using the radial kernel \eqref{eq:radial.kernel} with $D = 3$ and $\kappa = H/2\pi$, and $\epsilon=\ln 1.5$, we numerically obtain steady state distributions $\tilde{p}(h)$ for different Hubble values and the three different top masses. Fig.~\ref{fig:higgsdistributions} shows some of these distributions for the central value of top mass with different $H$, on a linear (upper panel) and log-log (lower panel) scale, respectively.

\begin{figure}[t!]
\centering
\includegraphics[width=1\columnwidth]{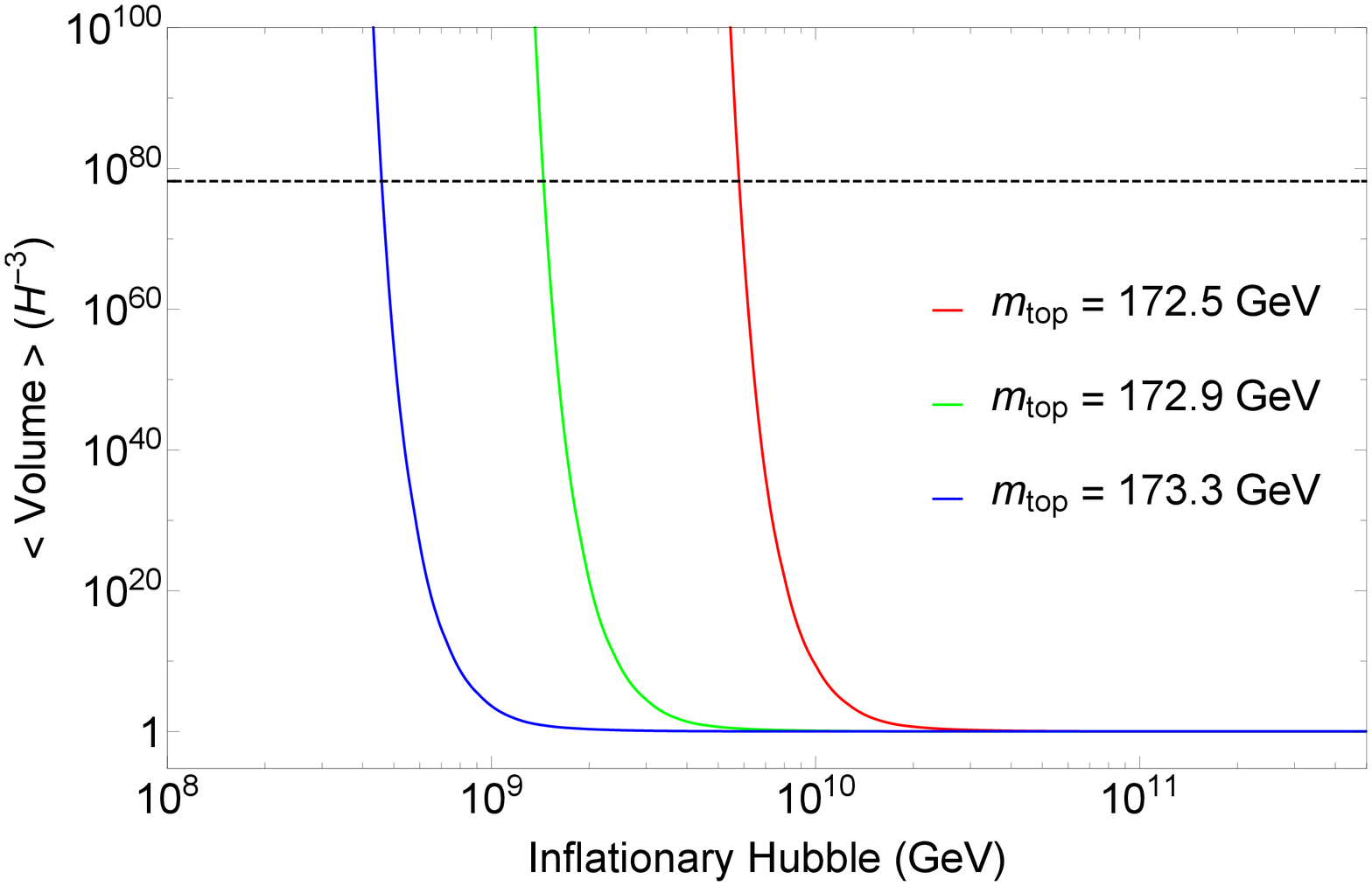}\vspace{0.5cm}\\
\includegraphics[width=1\columnwidth]{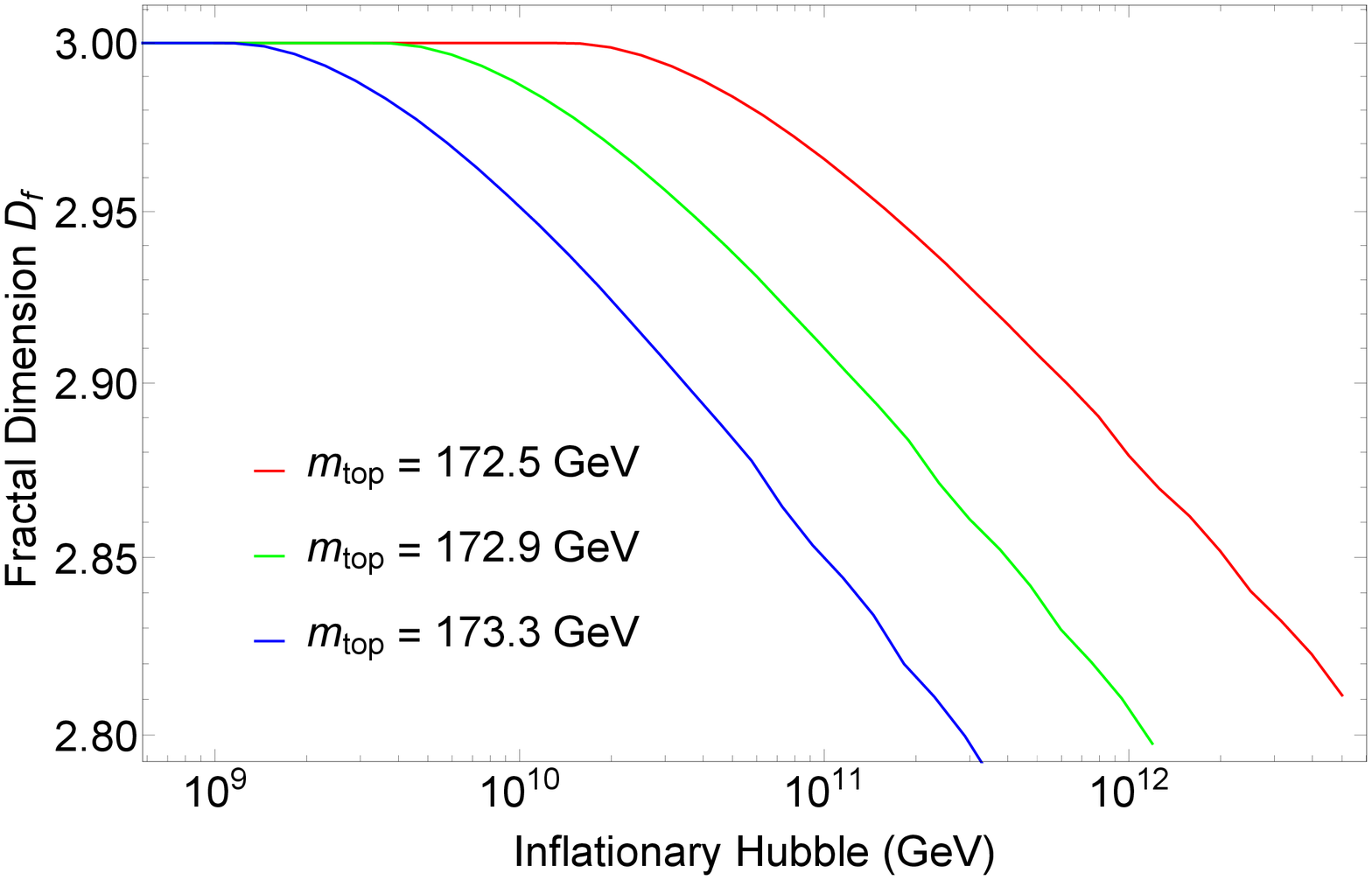}
\caption{\label{fig:averagevolume} Upper panel: Average volume of regions with Higgs within the instability scale in 3+1-dimensions within the minimal SM. 
Curves are obtained from steady state distributions.
The horizontal dashed black line corresponds to the characteristic size of our comoving universe $\sim e^{180}H^{-3}\approx10^{78}H^{-3}$ at the end of inflation. 
Lower panel: Fractal dimension of eternally inflating universe for regions with Higgs within the instability scale. This is within the framework of the minimal SM in 3+1-dimensions for 3 different values of the top mass.}
\end{figure}

We can use the kernel method then to again compute average volumes, trusting that the assumption of uncorrelated Hubble patches is not a terrible approximation. In general as a function of Hubble $H$, Fig.~\ref{fig:averagevolume} (upper panel) shows average volume of regions with $h < h_{\text{ht}}$ computed as before
\begin{eqnarray}
\langle\,\text{Volume}\,\rangle &=& \dfrac{1}{1-f_{h}}\,H^{-3};\nonumber\\
f_{h} &=& \int^{h_{\text{ht}}}_{0} d\,h\;\tilde{p}(h),
\label{eq:averagevolume}
\end{eqnarray}
for the three different values of top mass. We would like the average volume to be at least as big as our observable universe. Now note that we need $\approx 60$ e-foldings of inflation to produce our observable universe. This leads to a post-inflationary era made up of $\sim e^{180}$ inflationary Hubble patches. If we then expand that co-moving volume to today it will indeed be on the size of our observable universe. Hence the patches created in this steady state distribution should typically be at least this large, and ideally the average $<\mbox{Volume}>$ should be at least this large. We find that the bound on inflationary Hubble values, in order to achieve this is
\begin{eqnarray}
H_{\text{max}} =
\left\{
	\begin{array}{ll}
		5.8 \times 10^{9} \; \text{GeV} &  \mbox{for} \;\; m_{\text{top}} = 172.5 \; \text{GeV} \\
		1.4 \times 10^{9} \; \text{GeV}  &  \mbox{for} \;\; m_{\text{top}} = 172.9 \; \text{GeV} \\
		4.6 \times 10^{8} \; \text{GeV}  &  \mbox{for} \;\; m_{\text{top}} = 173.3 \; \text{GeV} \\
	\end{array}
\right.\,\,\,\,\,
\label{eq:Hmax}
\end{eqnarray}
which is one of our primary findings. 

We note that these values are less than the Gaussian approximated critical Hubble values $H^{(0)}_{\text{cr}}$. Hence one could in principle use the Gaussian approximation to examine steady state. However, in the literature, steady state is usually not utilized. Instead the distribution is usually simply cut off at $N\approx60$ to obtain a critical Hubble value. 


Finally, Fig.~\ref{fig:averagevolume} (lower panel) shows the fractal dimension $D_{f}$ of regions containing Higgs within the instability scale as a function of $H$, which is given as before, but with $D=3$ as the background dimension:
\begin{equation}
D_{f} = \lim_{N \rightarrow \infty}\dfrac{\ln\,V_{< h_{\text{ht}}}(N)}{N}
\end{equation}
with
\begin{eqnarray}
V_{< h_{\text{ht}}}(N) = e^{3\,\epsilon\,s} \prod^{s-1}_{i=0}\int^{h_{\text{ht}}}_{0}d\,h_{i}\,K\left(h_{i+1},h_{i},\epsilon\right)\,p\left(h_0,0\right)\nonumber.
\label{eq:greenvolume}
\end{eqnarray}
and $N = \epsilon\,s$.

As mentioned earlier, in any realistic scenario of our universe, after inflation it must reheat rendering thermal corrections extremely important in the analysis. Having dealt with the eternal inflation scenario at zero temperature, in the next section we will analyze these thermal corrections.

\section{Thermal Effects After Inflation}\label{Temperature.corrections}

It is known at finite temperature, the Higgs potential is partially improved, in the sense that at high temperatures the new effective potential has an instability scale $h_{\text{ht}}$ that is pushed to even higher values. In the literature it has sometimes been assumed that this can relax the bounds on the inflationary Hubble scale completely, i.e., any $H$ is now allowed, since even though the Higgs may be on the wrong side during inflation, it can be thrown back after inflation in the altered potential. Often authors assume ``instant" reheating, and some have used related ideas to build interesting models to explain various physical phenomenon (e.g. see~\cite{Espinosa.Riotto.2}). 

\begin{figure}[t!]
\centering
\includegraphics[width=1\columnwidth]{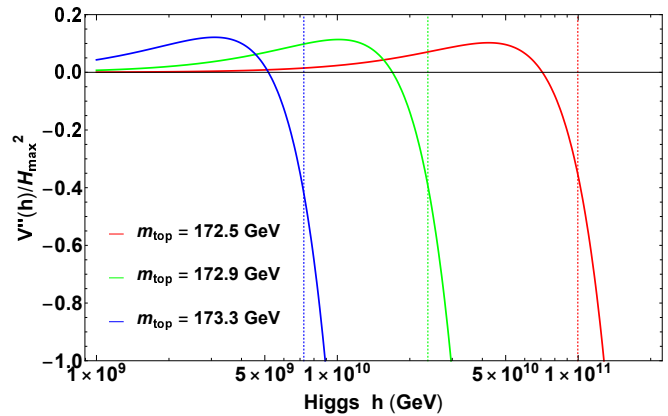}
\caption{\label{fig:FastRoll} A measure of the fast-roll $V''(h)/H^2$, with $H=H_{max}$ (the maximum $H$ value to allow for a large observable universe). The dashed vertical lines are $h=h_{\text{ht}}$ are the hill-top values for the Higgs. This plot shows that for $h>h_{\text{ht}}$ the field is about to undergo fast-roll and we expect it to readily head towards an AdS crunch or other catastrophe. This is within the framework of the minimal SM in 3+1-dimensions for 3 different values of the top mass.}
\end{figure}

However there are reasons to be skeptical that the Higgs field can be ``saved" during the reheating era. In Fig.~\ref{fig:FastRoll} we plot a measure of slow-roll, namely $V''/H^2$ (which is on the order of the second slow-roll parameter $\eta_{sr}$). This is the effective mass-squared in units of Hubble. This is evaluated at $H=H_{max}$ that we found above, and a similar curve applies for nearby value of $H$ too. The dashed lines are the turn-over scale $h_{\text{ht}}$ at zero temperature (during inflation). We see that for $h>h_{\text{ht}}$ then $|V''|/H_{Max}^2$ is appreciable, so the field is expected to start to roll-quickly. This gives very little time for reheating to occur and ``save" the Higgs field. For $H>H_{max}$, the field does not roll as fast, but this is an even more precarious situation since so much of the Higgs field is on the wrong side, and so it would be rather non-trivial for it all to be saved. In summary, it seems safe to assume that $H<H_{max}$ that we computed above. Although a full exploration of this issue is left for future work.

On the other hand, it was claimed in Ref.~\cite{Kohri.1} that even though $h_{\text{ht}}$ is pushed to higher values, thermal fluctuations in the Higgs can be so large and can throw the Higgs to the wrong side, even if $H<H_{max}$. This may be a particularly interesting approach when one is considering the formation of AdS regions where gravity will ultimately play a significant role. However, it is also of importance to note that no physical spatial scales were accounted for in this analysis or a direct inclusion of local gravitational effects. Thermal fluctuations were calculated at a \textit{point} in space, which was then used to construct a Gaussian probability distribution in the Higgs. This was then required to have small enough support for $h > h_{\text{ht}}$ such that a large  $\sim e^{180}H^{-3}$ sized region can survive, and a bound on the reheat temperature was found. However, this does not ensure an instability. The correct analysis is to consider the formation of thermal bubbles; these have finite spatial extent (roughly given by the inverse effective Higgs mass), and one needs to take into account the spatial dependence of the two-point correlation function. The appropriate calculation is the corresponding finite temperature Euclidean bounce solutions (instantons). Taking this into account, we will be led to different conclusions from Ref.~\cite{Kohri.1}. However, local gravitational effects could still alter the situation, and a full inclusion of such effects is an interesting topic. We also mention there is important earlier work on this thermal tunneling (including Refs.~\cite{EWvevmetastbl.3,EWvevmetastbl.4,Arnold:1991cv,Dine:1992wr,Delaunay:2007wb,Salvio:2016mvj}), which is connected to our work here.

We ignore the possibility of a direct coupling between the inflaton and the Higgs, leaving this for future work, and assume a generic reheating scenario that takes at least a few e-foldings. One may imagine reheating takes place by coupling the inflaton to gluons or other SM particles by higher dimension operators. Once the universe gets in thermal equilibrium with all the SM degrees of freedom, we can then attach the thermal free energy to the effective Higgs potential: The free energies due to bosons and fermions in the thermal bath are
\begin{eqnarray}
\Delta V_{B} (h,T) &=& \dfrac{T^4}{2\pi^2}\int^{\infty}_{0} dz\,z^2\ln\left(\dfrac{1 - e^{-\sqrt{z^2 + \frac{m_{B}^2(h)}{T^2}}}}{1-e^{-z}}\right)\nonumber\\
\Delta V_{F} (h,T) &=& \dfrac{-T^4}{2\pi^2}\int^{\infty}_{0} dz\, z^2\ln\left(\dfrac{1 + e^{-\sqrt{z^2 + \frac{m_{F}^2(h)}{T^2}}}}{1+e^{-z}}\right)\nonumber\\
\end{eqnarray}
respectively (we have subtracted out the radiation piece $\sim T^4$, since, although it is important to the total energy of the universe, it does not directly affect the Higgs dynamics as it is $h$-independent). The only significant contributions come from the W and Z bosons ($6$ and $3$ total degrees of freedom respectively), and the top quark ($12$ total degrees of freedom), each of which have the following masses
\begin{eqnarray}
m_{\text{W}}^2(h) &=& \dfrac{g^2(h)}{4}\,h^2\nonumber\\
m_{\text{Z}}^2(h) &=& \dfrac{g^2(h) + g'^2(h)}{4}\,h^2\nonumber\\
m_{\text{top}}^2(h) &=& \dfrac{y_t^2(h)}{2}\,h^2,
\end{eqnarray}
and where $g$, $g'$, and $y_t$ are the SU(2), U(1) and top Yukawa couplings respectively. The effective temperature corrected Higgs potential is therefore
\begin{eqnarray}
V_{\text{eff}}(h,T) &=& V(h,0) + \sum_{i = \text{W,Z,top}}\Delta V_{i} (h,T).
\label{eq:Vfull}
\end{eqnarray}
With this new effective potential, we can calculate bounce actions for tunneling of the Higgs to true vacuum at finite temperature. In the next section, we numerically obtain bounce actions for different temperatures with the above temperature corrected effective Higgs potential, and show that there is no bound on reheat temperature.

\subsection{Tunneling Probability at Finite Temperature}

\begin{figure}[t!]
\centering
\includegraphics[width=1\columnwidth]{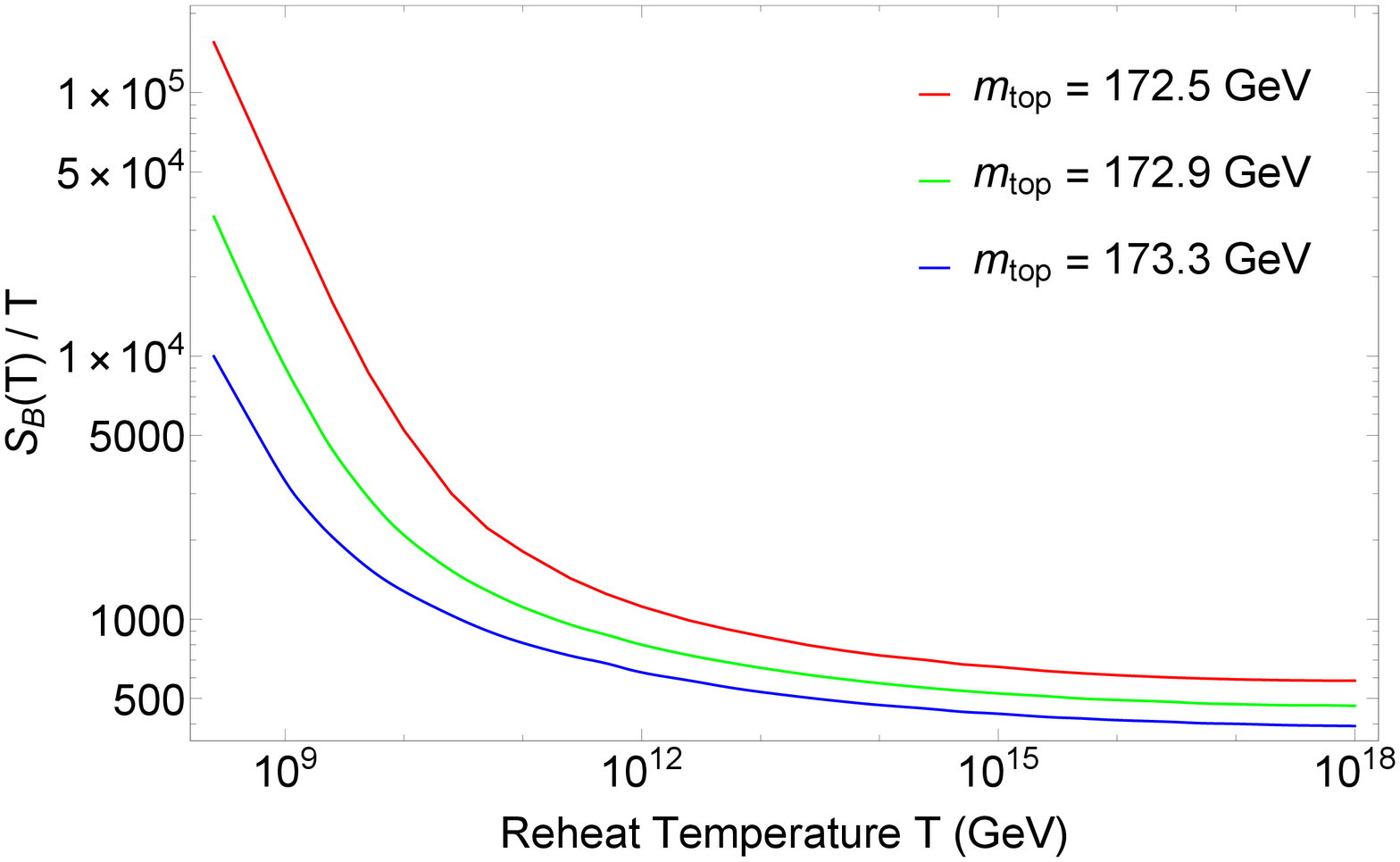}\vspace{0.5cm}\\
\includegraphics[width=1\columnwidth]{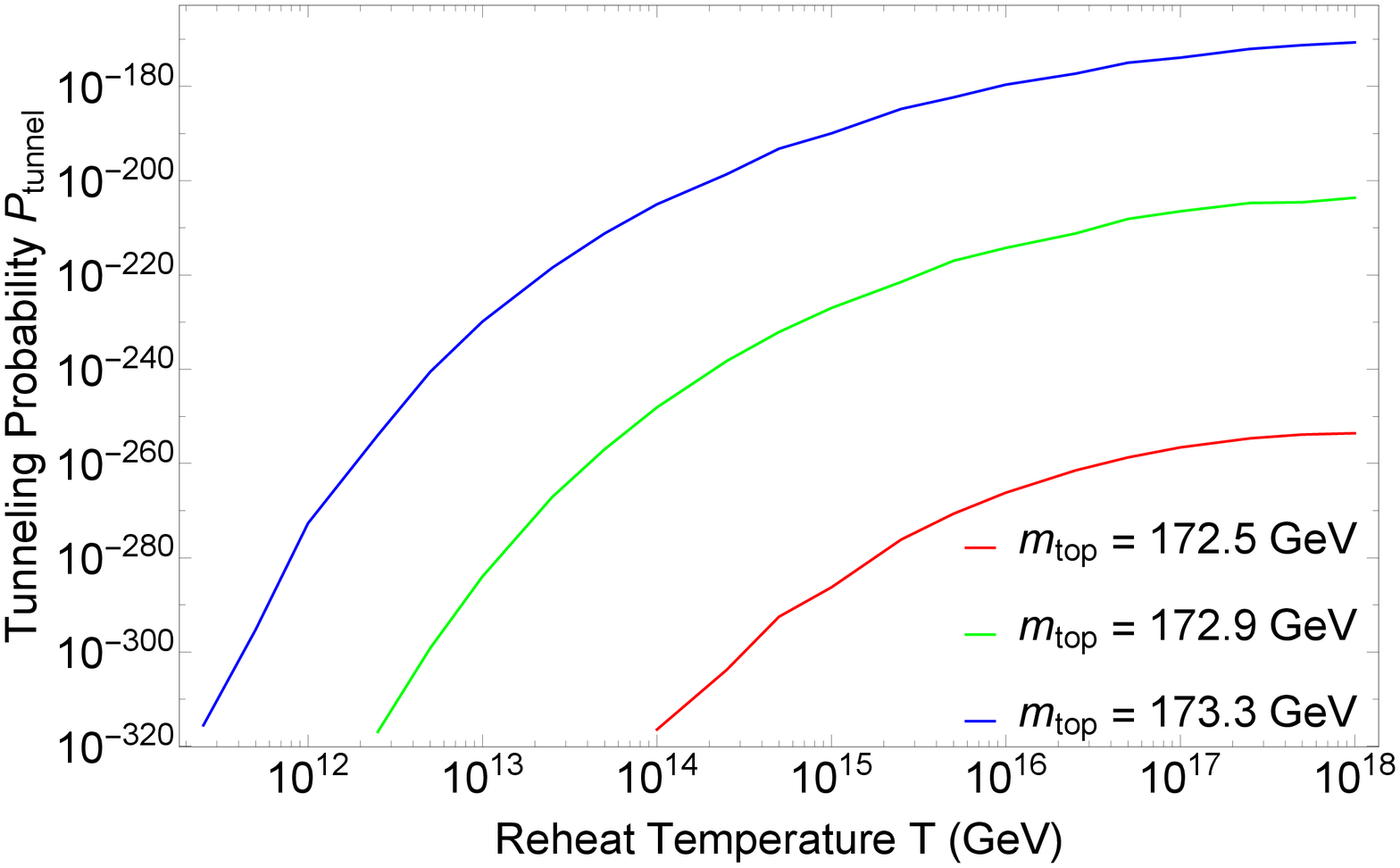}
\caption{\label{fig:tunnelprob} Upper panel: 3-dimensional Euclidean bounce action $S_{B}(T)/T$ vs reheat temperature $T$ for a thermal bubble. Lower panel: The corresponding tunneling probability $P_{\text{tunnel}} \sim e^{-S_{B}(T)/T}$. This is within the minimal SM and we have shown the result for 3 different top masses.}
\end{figure}

For a field theory at finite temperature in 3+1-dimensions, the relevant bounce action associated with a thermal bubble is provided by the Euclidean $O(3)$ solution~\cite{Linde.Bounce.temperature}
\begin{equation}
S_{B}(T) = 4\pi\int dr\,r^2\left[\dfrac{1}{2}\left(\dfrac{dh_{B}}{dr}\right)^2 + V(h_{B},T)\right]
\end{equation}
where $h_{B}(r,T)$ is the bounce solution for the equation of motion
\begin{equation}
\frac{d^2h}{dr^2} + \dfrac{2}{r}\dfrac{dh}{dr} = \dfrac{\partial\,V}{\partial\,r},
\end{equation}
with boundary conditions
\begin{eqnarray}
h(\infty,T) = 0,\nonumber\\
\dfrac{dh}{dr}(0,T) = 0.
\end{eqnarray}
Then, the probability of tunneling via thermal bubble formation is roughly (we suppress the pre-factor here as it is of sub-leading importance)
\begin{equation}
P_{\text{tunnel}} \propto \text{Exp}\left[-\dfrac{S_{B}(T)}{T}\right].
\end{equation}
With $V_{\text{eff}}(h,T)$, we numerically obtain bounce actions for different temperatures and hence tunneling probabilities. Fig.~\ref{fig:tunnelprob} (upper panel) shows $S_{B}(T)/T$ vs temperature $T$, and Fig~\ref{fig:tunnelprob} (lower panel) shows an estimate of corresponding tunneling probabilities by simply exponentiating the former. 

This probability, when properly normalized, is per unit volume per unit time. If we take the inverse of this probability, it is seen to be far greater than the number of bubble sized patches in our comoving universe (that is very roughly $\sim e^{180}\sim 10^{78}$, or so). It is clear that not only is there no bound on reheat temperature, even extremely high temperatures approaching Planck scale (though they would not realistically be reached in any post-inflationary era) carry insignificant tunneling probabilities for there to be a thermal Higgs bubble formation in our observable universe. 

\section{Summary}\label{Summary}

In this paper, we have for the first time analyzed the eternally inflating scenario in the context of the minimal SM. Since the SM Higgs field develops an instability at high field values, thereby allowing an inflationary phase only in a finite domain, the universe gets to an eternally inflating steady state in which any initial transient behavior washes out, and all interesting quantities like average volumes of regions, fractal dimensions, etc, are therefore independent of the number of e-foldings provided in the final stages of slow-roll inflation. 

The corresponding steady state distribution in the field has significant deviations from Gaussianity depending on the value of inflationary Hubble $H$ and is therefore important to keep track of. In order to obtain it for different choices of parameters, we derived a kernel that propagates the radial field's distribution in time, which upon normalization gives the desired steady state distribution. In order to compare it against Gaussian approximation, we began with a toy model of 4 scalar fields with a spherically symmetric M shaped potential in 1+1 dimensions and showed that the usual Gaussian approximated (radial) field distribution in any Hubble volume deviates significantly from the former. For a concrete analysis, we also compared them against actual 1+1-dimensional simulations, finding the kernel method to be much more accurate. We then used the radial field's kernel to analyze this eternal inflation scenario with the 2-loop renormalized SM Higgs potential. As before, we first compared kernel propagation analysis with actual 1+1-dimensional simulations, and showed that the former is accurate in calculating various statistical quantities. Then, we applied it to 3+1-dimensions and along with obtaining these statistical quantities, we obtained upper bounds on the inflationary Hubble scale $H$ to obtain an observable universe (of size $\sim e^{180} H^{-3}$) for upper, central, and lower value of top quark mass. 

Finally, after inflation ends locally in various regions, they must reheat and thermal corrections (due to W, Z and top quark) become important, leading to the instability scale of Higgs getting pushed to higher values for larger temperatures. Assuming a generic reheating scenario, where the inflaton takes at least a few Hubble times to reheat the universe, and therefore thermal effects are unlikely to rescue the Higgs if $H>H_{max}$, we then calculated tunneling probabilities from the electroweak vacuum at finite temperature to the true vacuum. We found that even for extremely high temperatures, the bounce actions are large enough that thermal bubble creation is negligibly rare in our comoving volume. Therefore, high reheating of the SM poses no threat to the electroweak vacuum metastability all the way to $M_{\text{pl}}$, but a high scale of inflation does. Modified reheating scenarios will be the subject of future work.

\vspace{-0.5cm}

\section*{Acknowledgments}
We would like to thank Alex Vilenkin for helpful discussion. 
MPH is supported in part by National Science Foundation grant PHY-1720332.

\vspace{0.5cm}

\appendix

\section{\label{appA}Beta functions}

Two loop beta functions for the relevant SM couplings and the anomalous dimension of Higgs $\gamma$, are as follows~\cite{Beta.ref.1,Beta.ref.2,Beta.ref.3,Beta.ref.4,Beta.ref.5}
\begin{widetext}
\begin{eqnarray}
\beta_{\lambda} &=& \dfrac{1}{16\,\pi^2}\left[24 \lambda^2 - 6 y_{t}^4 + \dfrac{3}{8}\left(2\,g^4 + (g^2 + g'^2)^2\right) + \lambda\left(12 y_t^2 - 9g^2 - 3g'^2\right)\right]\nonumber\\
&+& \left(\dfrac{1}{16\,\pi^2}\right)^2\left[30 y_t^6 - y_t^4\left(32\,g_s^2 + \dfrac{8}{3}g'^2 + 3\lambda\right) + y_t^2\left(-\dfrac{9}{4}g^4 + \dfrac{21}{2}g^2\,g'^2 - \dfrac{19}{4}\,g'^4 + \lambda\left(80\,g_s^2 + \dfrac{45}{2}g^2 + \dfrac{85}{6}g'^2 - 144\,\lambda\right)\right)\right.\nonumber\\
&+&\left. \dfrac{1}{48}\left(915\,g^6 - 289\,g^4\,g'^2 - 559\,g^2\,g'^4 - 379\,g'^6\right) + \lambda\left(-\dfrac{73}{8}g^4 + \dfrac{39}{4}g^2\,g'^{\,2} + \dfrac{629}{24}\,g'^{\,4} + 108\,\lambda\,g^2 + 36\,\lambda\,g'^{\,2} - 312\,\lambda^2\right)\right];\nonumber\\
\beta_{y_t} &=& \dfrac{y_t}{16\,\pi^2}\left[\dfrac{9}{2}y_t^2 - \left(\dfrac{17}{12}g'^{\,^2} + \dfrac{9}{4}g^2 + 8\,g_s^2\right)\right] + y_t\left(\dfrac{1}{16\,\pi^2}\right)^2\left[6\left(\lambda^2 - 2\,y_t^4 - 2\,\lambda\,y_t^2\right) + y_t^2\left(\dfrac{131}{16}g'^{\,2} + \dfrac{225}{16}g^2 + 36\,g_s^2\right)\right.\nonumber\\
&+& \left.\dfrac{1187}{216}g'^{\,4} - \dfrac{23}{4}g^4 - \dfrac{3}{4}g'^{\,2}\,g^2 - 108\,g_s^4 + 9\,g_s^2\,g^2 + \dfrac{19}{9}g_s^2\,g'^{\,2}\right];\nonumber\\
\beta_{g} &=& \dfrac{1}{16\,\pi^2}\left[\dfrac{10}{3} - \dfrac{1}{6}\right]g^3 + \left(\dfrac{1}{16\,\pi^2}\right)^2g^3\left[\dfrac{35}{6}g^2 + \dfrac{3}{2}g'^{\,2} + 12\,g_s^2 - \dfrac{3}{2}y_t^2\right];\nonumber\\
\beta_{g'} &=& \dfrac{1}{16\,\pi^2}\left[\dfrac{20}{3} + \dfrac{1}{6}\right]g'^{\,3} + \left(\dfrac{1}{16\,\pi^2}\right)^2g^{\,3}\left[\dfrac{9}{2}g^2 + \dfrac{199}{18}g'^{\,2} + \dfrac{44}{3}g_s^2 - \dfrac{17}{6}y_t^2\right];\nonumber\\
\beta_{g_s} &=& \dfrac{1}{16\,\pi^2}7\,g_s^3 + \left(\dfrac{1}{16\,\pi^2}\right)^2g_s^3\left[\dfrac{9}{2}g^2 + \dfrac{11}{6}g'^{\,2} - 26\,g_s^2 - 2\,y_t^2\right];\nonumber\\
\gamma &=& -\dfrac{1}{16\,\pi^2}\left[\dfrac{9}{4}g^2 + \dfrac{3}{4}g'^{\,2} - 3\,y_t^2\right] - \left(\dfrac{1}{16\,\pi^2}\right)^2\left[\dfrac{27}{4}y_t^4 - \dfrac{5}{2}y_t^2\left(8\,g_s^2 + \dfrac{9}{4}g^2 + \dfrac{17}{12}g'^{\,2}\right) + \dfrac{271}{32}g^4 - \dfrac{9}{16}g^2\,g'^{\,2} - \dfrac{431}{96}g'^{\,4} - 6\,\lambda^2\right].\nonumber
\label{eq:beta.functions}
\end{eqnarray}
\end{widetext}

Here, $\lambda$, $y_{t}$, $g$, $g'$, and $g_s$ are the Higgs self-coupling, top quark Yukawa coupling, weak SU(2) coupling, U(1) hyper-charge coupling, and strong coupling respectively.


\begin{thebibliography}{10}

\bibitem{MM.1} 
  M.~Jain and M.~P.~Hertzberg,
  ``Statistics of Inflating Regions in Eternal Inflation,''
  Phys.\ Rev.\ D {\bf 100}, no. 2, 023513 (2019)
  [arXiv:1904.04262 [astro-ph.CO]].
  %
\bibitem{Higgs.discovery}
The ATLAS Collaboration,
"Observation of a new particle in the search for the Standard Model Higgs boson with the ATLAS detector at the LHC,"
Phys. Lett. {\bf B716} (2012) 1-29
%
\bibitem{Particle.data.latest}
M.~Tanabashi et al. (Particle Data Group), 
"Review of Particle Physics",
Phys. Rev. D {\bf 98}, 030001 (2018).
%
\bibitem{EWvevmetastbl.1}
G.~Degrassi et al., 
"Higgs mass and vacuum stability in the Standard Model at NNLO",
JHEP {\bf 08} (2012) 098.
%
\bibitem{EWvevmetastbl.2}
G.~Isidori, G.~Ridolfi and A.~Strumia, 
"On the metastability of the Standard Model vacuum",
Nucl. Phys. {\bf B 609} (2001) 387.
%
\bibitem{EWvevmetastbl.3}
J.~Ellis, J.~R.~Espinosa, G.~F.~Giudice, A.~Hoecker and A.~Riotto, 
"The probable fate of the Standard Model",
Phys. Lett. {\bf B 679} (2009) 369.
%
\bibitem{EWvevmetastbl.4}
J.~Elias-Miro, J.~R.~Espinosa, G.~F.~Giudice, G.~Isidori, A.~Riotto and A.~Strumia, 
"Higgs mass implications on the stability of the electroweak vacuum",
Phys. Lett. {\bf B 709} (2012) 222.
%
\bibitem{A.Guth}
A.~H.~Guth
"The inflationary universe: A possible solution to the horizon and flatness problems" 
Phys. Rev. D {\bf 23}, 347 (1981)
%
\bibitem{A.Linde.1}
A.~D.~Linde
"A new inflationary universe scenario: A possible solution of the horizon, flatness, homogeneity, isotropy and primordial monopole problems"
Phys. Lett. {\bf 108B}, 389 (1982)
%
\bibitem{A.Linde.2}
A.~D.~Linde, 
"Chaotic inflation",
Phys. Lett. {\bf 129B}, 177 (1983).
%
\bibitem{A.Vilenkin}
A.~Vilenkin, 
"The birth of inflationary universes",
Phys. Rev. D {\bf 27}, 2848 (1983).
%
\bibitem{Espinosa.Riotto.1}
J.~R.~Espinosa, G.~F.~Giudice, E.~Morgante, A.~Riotto, L.~Senatore, A.~Strumia, and N.~Tetradis, 
"The cosmological Higgstory of the vacuum instability",
 J. High Energy Phys. {\bf 2015}, 174.
%
\bibitem{Zurek.1}
J.~Kearney, H.~Yoo, and K.~M.~Zurek, 
"Is a Higgs vacuum instability fatal for high-scale inflation?",
 Phys. Rev. D {\bf 91}, 123537 (2015).
%
\bibitem{Zurek.2}
W.~E.~East, J.~Kearney, B.~Shakya, H.~Yoo, and K.~M.~Zurek, 
"Spacetime dynamics of a Higgs vacuum instability during inflation",
Phys. Rev. D {\bf 95}, 023526 (2017).
%
\bibitem{Kohri.1}
K. Kohri and H. Matsui, 
"Higgs vacuum metastability in primordial inflation, preheating, and reheating",
 Phys. Rev. D {\bf 94}, 103509 (2016).
%
\bibitem{Kohri.2}
K. Kohri and H. Matsui, 
"Electroweak Vacuum Instability and Renormalized Higgs Field Vacuum Fluctuations in the Inflationary Universe",
JCAP 08 (2017) {\bf 011}.
%
\bibitem{Espinosa.Riotto.2}
J.~R.~Espinosa, D.~Racco, and A.~Riotto, 
"Cosmological Signature of the Standard Model Higgs Vacuum Instability: Primordial Black Holes as Dark Matter",
Phys. Rev. Lett. {\bf 120}, 121301.
%
 \bibitem{Figueroa:2015rqa} 
  D.~G.~Figueroa, J.~Garcia-Bellido and F.~Torrenti,
  ``Decay of the standard model Higgs field after inflation,''
  Phys.\ Rev.\ D {\bf 92}, no. 8, 083511 (2015)
  [arXiv:1504.04600 [astro-ph.CO]].
  %
  \bibitem{Ema:2016kpf} 
  Y.~Ema, K.~Mukaida and K.~Nakayama,
  ``Fate of Electroweak Vacuum during Preheating,''
  JCAP {\bf 1610}, 043 (2016)
  [arXiv:1602.00483 [hep-ph]].
  %
  \bibitem{Enqvist:2016mqj} 
  K.~Enqvist, M.~Karciauskas, O.~Lebedev, S.~Rusak and M.~Zatta,
  ``Postinflationary vacuum instability and Higgs-inflaton couplings,''
  JCAP {\bf 1611}, 025 (2016)
  [arXiv:1608.08848 [hep-ph]].
   %
  \bibitem{Gross:2015bea} 
  C.~Gross, O.~Lebedev and M.~Zatta,
  ``Higgs–inflaton coupling from reheating and the metastable Universe,''
  Phys.\ Lett.\ B {\bf 753}, 178 (2016)
  [arXiv:1506.05106 [hep-ph]].
  %
  \bibitem{Espinosa:2018eve} 
  J.~R.~Espinosa, D.~Racco and A.~Riotto,
  ``A Cosmological Signature of the SM Higgs Instability: Gravitational Waves,''
  JCAP {\bf 1809}, 012 (2018)
  [arXiv:1804.07732 [hep-ph]].
  %
  \bibitem{Joti:2017fwe} 
  A.~Joti, A.~Katsis, D.~Loupas, A.~Salvio, A.~Strumia, N.~Tetradis and A.~Urbano,
  ``(Higgs) vacuum decay during inflation,''
  JHEP {\bf 1707}, 058 (2017)
  [arXiv:1706.00792 [hep-ph]].
  %
  \bibitem{Ema:2016ehh} 
  Y.~Ema, K.~Mukaida and K.~Nakayama,
  ``Electroweak Vacuum Stabilized by Moduli during/after Inflation,''
  Phys.\ Lett.\ B {\bf 761}, 419 (2016)
  [arXiv:1605.07342 [hep-ph]].
  %
  \bibitem{Markkanen:2018pdo} 
  T.~Markkanen, A.~Rajantie and S.~Stopyra,
  ``Cosmological Aspects of Higgs Vacuum Metastability,''
  Front.\ Astron.\ Space Sci.\  {\bf 5}, 40 (2018)
  [arXiv:1809.06923 [astro-ph.CO]].
  %
  \bibitem{Saha:2016ozn} 
  A.~K.~Saha and A.~Sil,
  ``Higgs Vacuum Stability and Modified Chaotic Inflation,''
  Phys.\ Lett.\ B {\bf 765}, 244 (2017)
  [arXiv:1608.04919 [hep-ph]].
  %
  \bibitem{Espinosa:2018euj} 
  J.~R.~Espinosa, D.~Racco and A.~Riotto,
  ``Primordial Black Holes from Higgs Vacuum Instability: Avoiding Fine-tuning through an Ultraviolet Safe Mechanism,''
  Eur.\ Phys.\ J.\ C {\bf 78}, no. 10, 806 (2018)
  [arXiv:1804.07731 [hep-ph]].
  %
  \bibitem{Ema:2017rkk} 
  Y.~Ema, K.~Mukaida and K.~Nakayama,
  ``Electroweak Vacuum Metastability and Low-scale Inflation,''
  JCAP {\bf 1712}, 030 (2017)
  [arXiv:1706.08920 [hep-ph]].
  %
  \bibitem{Gong:2017mwt} 
  J.~O.~Gong and N.~Kitajima,
  ``Cosmological stochastic Higgs field stabilization,''
  Phys.\ Rev.\ D {\bf 96}, no. 6, 063521 (2017)
  [arXiv:1705.11178 [hep-ph]].
%
\bibitem{A.Linde.3}
A.~D.~Linde and A.~Mezhlumian, 
"Stationary universe",
Phys. Lett. B {\bf 307}, 25 (1993).
%
\bibitem{A.Linde.4}
A.~D.~Linde, D.~A.~Linde, and A.~Mezhlumian, 
"From the Big Bang theory to the theory of a stationary universe",
Phys. Rev. D {\bf 49}, 1783 (1994).
%
\bibitem{Hertzberg:2018kyi} 
  M.~P.~Hertzberg and M.~Jain,
  ``Counting of States in Higgs Theories,''
  Phys.\ Rev.\ D {\bf 99}, no. 6, 065015 (2019)
  [arXiv:1807.05233 [hep-th]].
%
  \bibitem{EliasMiro:2012ay} 
  J.~Elias-Miro, J.~R.~Espinosa, G.~F.~Giudice, H.~M.~Lee and A.~Strumia,
  ``Stabilization of the Electroweak Vacuum by a Scalar Threshold Effect,''
  JHEP {\bf 1206}, 031 (2012)
  [arXiv:1203.0237 [hep-ph]].
%
\bibitem{Hertzberg:2012zc} 
  M.~P.~Hertzberg,
  ``A Correlation Between the Higgs Mass and Dark Matter,''
  Adv.\ High Energy Phys.\  {\bf 2017}, 6295927 (2017)
  [arXiv:1210.3624 [hep-ph]].
  %
  \bibitem{Arnold:1991cv}
P.~B.~Arnold and S.~Vokos,
``Instability of hot electroweak theory: bounds on m(H) and M(t),''
Phys.\ Rev.\ D \textbf{44}, 3620-3627 (1991)
 %
  \bibitem{Dine:1992wr} 
  M.~Dine, R.~G.~Leigh, P.~Y.~Huet, A.~D.~Linde and D.~A.~Linde,
  ``Towards the theory of the electroweak phase transition,''
  Phys.\ Rev.\ D {\bf 46}, 550 (1992)
  [hep-ph/9203203].
  %
  \bibitem{Delaunay:2007wb} 
  C.~Delaunay, C.~Grojean and J.~D.~Wells,
  ``Dynamics of Non-renormalizable Electroweak Symmetry Breaking,''
  JHEP {\bf 0804}, 029 (2008)
  [arXiv:0711.2511 [hep-ph]].
 %
 \bibitem{Salvio:2016mvj} 
 A.~Salvio, A.~Strumia, N.~Tetradis and A.~Urbano,
 ``On gravitational and thermal corrections to vacuum decay,''
 JHEP {\bf 1609}, 054 (2016)
 [arXiv:1608.02555 [hep-ph]].
 %
\bibitem{Linde.Bounce.temperature} 
Andrei Linde,
``Stochastic approach to tunneling and baby universe formation,''
Nucl. Phys. {\bf B372}, 421-442 (1992).
%
\bibitem{Beta.ref.1}
M.~B.~Einhorn and D.~R.~T.~Jones, 
"Effective potential and quadratic divergences",
Phys. Rev. D {\bf 46}, 5206 (1992).
%
\bibitem{Beta.ref.2}
M.~E.~Machacek and M.~T.~Vaughn, 
"Two Loop Renormalization Group Equations in a General Quantum Field Theory. 1. Wave Function Renormalization",
Nucl. Phys. {\bf B222}, 83 (1983).
%
\bibitem{Beta.ref.3}
M.~E.~Machacek and M.~T.~Vaughn,
"Two-loop renormalization group equations in a general quantum field theory (II). Yukawa couplings",
Nucl. Phys. {\bf B236}, 221 (1984).
%
\bibitem{Beta.ref.4}
M.~E.~Machacek and M.~T.~Vaughn,
"Two Loop Renormalization Group Equations in a General Quantum Field Theory. 3. Scalar Quartic Couplings",
Nucl. Phys. {\bf B249}, 70 (1985).
%
\bibitem{Beta.ref.5}
C.~Ford, I.~Jack, and D.~R.~T.~Jones, 
"The Standard model effective potential at two loops",
 Nucl. Phys. {\bf B387}, 373 (1992); {\bf B504}, 551(E) (1997).
 


  
  
\end{thebibliography}
\end{document}